\documentclass[12pt,english]{article}
\usepackage{times}
\usepackage[T1]{fontenc}
\usepackage{geometry}
\usepackage{rotating}
\usepackage{graphicx}
\usepackage{setspace}
\usepackage{amssymb}

\makeatletter

\providecommand{\tabularnewline}{\\}

\usepackage{bm}

\usepackage{babel}
\makeatother
\begin{document}

\section*{\noindent Sensitivity Analysis for Antenna Devices through Interval
Arithmetic - A Generalized Approach}

\noindent ~

\noindent \vfill

\noindent N. Anselmi,$^{(1)(2)}$ \emph{Senior Member, IEEE}, P. Rocca,$^{(1)(2)(3)}$
\emph{Fellow, IEEE}, \emph{}and A. Massa,$^{(1)(2)(4)(5)(6)}$ \emph{Fellow,
IEEE}

\noindent \vfill

\noindent {\footnotesize ~}{\footnotesize \par}

\noindent {\scriptsize $^{(1)}$} \emph{\scriptsize ELEDIA Research
Center} {\scriptsize (}\emph{\scriptsize ELEDIA}{\scriptsize @}\emph{\scriptsize UniTN}
{\scriptsize - University of Trento)}{\scriptsize \par}

\noindent {\scriptsize DICAM - Department of Civil, Environmental,
and Mechanical Engineering}{\scriptsize \par}

\noindent {\scriptsize Via Mesiano 77, 38123 Trento - Italy}{\scriptsize \par}

\noindent \textit{\emph{\scriptsize E-mail:}} {\scriptsize \{}\emph{\scriptsize nicola.anselmi.1,
paolo.rocca, andrea.massa}{\scriptsize \}@}\emph{\scriptsize unitn.it}{\scriptsize \par}

\noindent {\scriptsize Website:} \emph{\scriptsize www.eledia.org/eledia-unitn}{\scriptsize \par}

\noindent {\scriptsize ~}{\scriptsize \par}

\noindent {\scriptsize $^{(2)}$} \emph{\scriptsize CNIT - \char`\"{}University
of Trento\char`\"{} ELEDIA Research Unit }{\scriptsize \par}

\noindent {\scriptsize Via Sommarive 9, 38123 Trento - Italy}{\scriptsize \par}

\noindent {\scriptsize Website:} \emph{\scriptsize www.eledia.org/eledia-unitn}{\scriptsize \par}

\noindent {\scriptsize ~}{\scriptsize \par}

\noindent {\scriptsize $^{(3)}$} \emph{\scriptsize ELEDIA Research
Center} {\scriptsize (}\emph{\scriptsize ELEDIA}{\scriptsize @}\emph{\scriptsize XIDIAN}
{\scriptsize - Xidian University)}{\scriptsize \par}

\noindent {\scriptsize P.O. Box 191, No.2 South Tabai Road, 710071
Xi'an, Shaanxi Province - China }{\scriptsize \par}

\noindent {\scriptsize E-mail:} \emph{\scriptsize paolo.rocca@xidian.edu.cn}{\scriptsize \par}

\noindent {\scriptsize Website:} \emph{\scriptsize www.eledia.org/eledia-xidian}{\scriptsize \par}

\noindent {\scriptsize ~}{\scriptsize \par}

\noindent {\scriptsize $^{(4)}$} \emph{\scriptsize ELEDIA Research
Center} {\scriptsize (}\emph{\scriptsize ELEDIA}{\scriptsize @}\emph{\scriptsize UESTC}
{\scriptsize - UESTC)}{\scriptsize \par}

\noindent {\scriptsize School of Electronic Science and Engineering,
Chengdu 611731 - China}{\scriptsize \par}

\noindent \textit{\emph{\scriptsize E-mail:}} \emph{\scriptsize andrea.massa@uestc.edu.cn}{\scriptsize \par}

\noindent {\scriptsize Website:} \emph{\scriptsize www.eledia.org/eledia}{\scriptsize -}\emph{\scriptsize uestc}{\scriptsize \par}

\noindent {\scriptsize ~}{\scriptsize \par}

\noindent {\scriptsize $^{(5)}$} \emph{\scriptsize ELEDIA Research
Center} {\scriptsize (}\emph{\scriptsize ELEDIA@TSINGHUA} {\scriptsize -
Tsinghua University)}{\scriptsize \par}

\noindent {\scriptsize 30 Shuangqing Rd, 100084 Haidian, Beijing -
China}{\scriptsize \par}

\noindent {\scriptsize E-mail:} \emph{\scriptsize andrea.massa@tsinghua.edu.cn}{\scriptsize \par}

\noindent {\scriptsize Website:} \emph{\scriptsize www.eledia.org/eledia-tsinghua}{\scriptsize \par}

\noindent {\footnotesize ~}{\footnotesize \par}

\noindent {\footnotesize $^{(6)}$} {\scriptsize School of Electrical
Engineering}{\scriptsize \par}

\noindent {\scriptsize Tel Aviv University, Tel Aviv 69978 - Israel}{\scriptsize \par}

\noindent \textit{\emph{\scriptsize E-mail:}} \emph{\scriptsize andrea.massa@eng.tau.ac.il}{\scriptsize \par}

\noindent {\scriptsize Website:} \emph{\scriptsize https://engineering.tau.ac.il/}{\scriptsize \par}

~

\emph{This work has been submitted to the IEEE for possible publication.
Copyright may be transferred withoutnotice, after which this version
may no longer be accessible.}

\newpage
\section*{Sensitivity Analysis for Antenna Devices through Interval Arithmetic
- A Generalized Approach}

~

\noindent ~

\noindent ~

\begin{flushleft}N. Anselmi, P. Rocca, and A. Massa\end{flushleft}

\vfill

\begin{abstract}
\noindent This paper presents a novel method for the sensitivity analysis
of electromagnetic (\emph{EM}) systems whose transfer function (\emph{TF}),
that is the input-output (I/O) relationship between the input parameters
affected by tolerance and the system response (i.e., the arising \emph{EM}
performance of interest), is not available in closed (or explicit)
form. The method is a generalized analytic technique based on the
Interval Analysis (\emph{IA}). First, an analytic surrogate model
(\emph{SM}) of the \emph{TF} is defined by means of a learning-by-example
(\emph{LBE}) approach starting from a set of available I/O examples.
Then, the \emph{LBE}-derived \emph{SM} is extended to intervals through
\emph{IA} to yield inclusive, yet finite, performance bounds of the
output system response when the control parameters are affected by
unknown, but bounded, tolerances. A set of representative numerical
examples is reported to validate the proposed \emph{IA-LBE} method
as well as to assess its effectiveness and reliability when dealing
with realistic \emph{EM} systems (e.g., antennas) for which the \emph{TF}
is not explicitly known.

\noindent \vfill
\end{abstract}
\noindent \textbf{Key words}: Electromagnetics, Antennas, Antenna
Arrays, Interval Analysis, Learning-by-Examples, Sensitivity Analysis,
Surrogate Model, Tolerances.

\newpage
\section{Introduction\label{sec:Introduction}}

\noindent The analysis of the sensitivity of the performance of a
device on the deviation from the nominal values of its descriptors
is becoming more and more crucial and, at the same time, more and
more complex in electromagnetic (\emph{EM}) engineering due to the
growing proliferation of wireless technologies in nowadays applications
and their use in increasingly challenging conditions (e.g., harsh
environments, high operation frequencies, etc ...). Indeed, wireless
and antenna systems are sensitive to environmental factors, such as
the temperature and the humidity, as well as the presence of nearby
objects that can alter their control parameters and, thus, their \emph{EM}
response. Tolerances and small variations are also unavoidable in
any manufacturing process. Since wireless systems, including antennas
and other \emph{EM} devices, are designed to guarantee the fulfillment
of user-defined radiation requirements (e.g., power pattern masks
or thresholds on the gain and/or efficiency), sensitivity analysis
tools allow engineers to quantify how the variations on the control
parameters affect the system performance so that it is possible to
implement robust mitigation strategies. Doing so, the design turns
out to be more resilient, during the system operation, to the possible
deviations from the nominal values of the control parameters. Moreover,
the knowledge of the most critical - in terms of system performance
- parameters of an \emph{EM} device gives the opportunity to roll
out appropriate policies for assuring its robustness such as, for
instance, the use of more expensive, but more reliable, components
or considering a redundancy in some part of the \emph{EM} system.

\noindent As for antenna systems, several sensitivity analysis techniques
have been proposed to predict the effects of deviations from the nominal
values of the design parameters. In the state-of-the-art literature,
statistical approaches have been firstly introduced to provide simple
closed-form expressions for the mean value and the variance of the
power pattern and other radiation features (e.g., the main-lobe peak,
the sidelobe level (\emph{SLL}), and the pattern nulls) \cite{Ruze 1952}-\cite{Haupt 2010}.
Although very efficient and simple to apply, these methods are based
on the central limit theorem, thus they are applicable only under
small error conditions. More accurate predictions can be obtained
thanks to intensive numerical simulations in Monte Carlo (\emph{MC})
techniques \cite{Lee 2005}\cite{Hu 2022}. These latter do not consider
approximations and enable the sensitivity analysis of realistic \emph{EM}
devices, which are carefully modeled with accurate full-wave simulators.
However, owing to the high computational burden and the need to run
a number of simulations rapidly growing with the number of parameters,
their use is practically unfeasible for nowadays antenna systems often
featuring complex unconventional structures and multi-scale materials.
To reduce the computational burden of \emph{MC} techniques, other
approaches have been proposed, among which noteworthy are the Taylor-approximation
method \cite{Hu 2017}, the optimization-based worst-case analysis
\cite{Zhang 2018}, the stochastic method-of-moment (\emph{MoM}) \cite{Hatamkhani 2022},
and the polynomial chaos expansion \cite{Du 2017}\cite{Tomy 2019}.
More recently, thanks to the development of Learning-by-Examples (\emph{LBE})
or Machine Learning (\emph{ML}) techniques \cite{Massa 2018}, statistical
surrogate models, which infer the mapping between input variables
and output objectives from a set of known I/O examples, have been
exploited for their computational efficiency and advantages with respect
to full-wave \emph{EM} predictions \cite{Easum 2018}.

\noindent Starting from a different perspective from that of previous
methodologies, sensitivity analysis methods based on the Interval
Analysis (\emph{IA}) have been widely developed in the last decade
to efficiently compute worst-case tolerance bounds of the \emph{EM}
system performance. By knowing the maximum deviation from the nominal
value of the input parameters, analytic expressions for the bounds
of the \emph{EM} response have been derived by exploiting the arithmetic
of intervals \cite{Moore 1966}-\cite{Moore 2009}. The key advantage
of an \emph{IA}-based sensitivity analysis is that the computed bounds
are finite and reliable, since they include all possible realizations
of the I/O process by virtue of the \emph{IA} inclusion theorem \cite{Moore 1966}.
Moreover, the computation of the bounds is analytic, so computationally
very efficient, and the only terms involved in interval arithmetic
steps are the endpoints of the intervals of the input parameters.
Thanks to these positive features, \emph{IA} has been successfully
used for the tolerance analysis of several types of \emph{EM} devices
and systems, including array antennas \cite{Anselmi 2013}-\cite{Rocca 2020},
reflector antennas \cite{Rocca 2014a}\cite{Rocca 2014b}, reflectarrays,
\cite{Rocca 2016}, inflatable antennas \cite{Wang 2023}, and radomes
\cite{Li 2017a}-\cite{Xu 2023}. Up to now, the main drawback of
\emph{IA}-based sensitivity techniques was the need for explicit (or
closed-form) knowledge of the relationship between the antenna parameters,
affected by the uncertainty, and the \emph{EM} system response (e.g.,
the radiation pattern). Otherwise, it was not possible to apply these
methods.

\noindent This work is aimed at overcoming such a limit to extend
the application of \emph{IA}-based methods to \emph{EM} systems and
devices (e.g., antennas) where the relationship between the input
parameters and the \emph{EM} response in not \emph{a-priori} know,
but only in an \emph{implicit} form thanks to the availability of
a suitable number of I/O pairs/examples. More in detail, starting
from a set of available system responses for given input parameter
values, a generalized analytic surrogate model (\emph{SM}) of the
transfer function (\emph{TF}) of the \emph{EM} system is defined thanks
to a \emph{LBE} approach \cite{Massa 2018}. The \emph{IA} is then
used to extend to intervals the analytic expression of the \emph{LBE}-generated
\emph{SM}, thus determining the bounds of the \emph{EM} system response.
To the best of the authors' knowledge, the main novelties of this
work over the existing state-of-the-art literature include (\emph{i})
the introduction of an innovative \emph{IA}-based methodology to give
inclusive, yet finite, performance bounds for the sensitivity analysis
of arbitrary \emph{EM} devices whose control parameters are affected
by unknown, but bounded, tolerances, (\emph{ii}) the development of
a generalized analytic procedure, based on a \emph{LBE} technique,
for the derivation of a surrogate model of the \emph{EM} device response
to enable the use of \emph{IA}, and \emph{}(\emph{iii}) the derivation
of guidelines for the choice of the \emph{LBE} training set in order
to yield reliable, but non trivial (i.e., finite), \emph{IA} bounds.

\noindent The rest of the paper is organized as follows. The sensitivity
analysis problem at hand is formulated in Sect. 2, while the proposed
\emph{IA-LBE} method is presented in Sect. 3. Section 4 is devoted
to the numerical analysis and the assessment of the proposed approach.
Eventually, some conclusions and final remarks are drawn (Sect. 5).

\section{\noindent Mathematical Formulation\label{sec:Mathematical-Formulation}}

Without loss of generality, let us assume time-harmonic conditions
by omitting hereinafter, for the sake of presentation simplicity,
the time dependency factor $exp\left(j\omega t\right)$, $j$ , $\omega$,
and $t$ being the complex variable ($j=\sqrt{-1}$), the angular
frequency, and the time variable, respectively. Moreover, let $\mathbf{J}\left(\mathbf{r}\,;\underline{p}\right)$
be the representative function of a generic \emph{EM} device (e.g.,
the current density distribution of an antenna),\begin{equation}
\mathbf{J}\left(\mathbf{r}\,;\underline{p}\right)=\sum_{i=\left\{ x,y,z\right\} }J_{i}\left(\mathbf{r};\underline{p}\right)\hat{\mathbf{u}}_{i},\label{eq:_current.LBE-input}\end{equation}
$\mathbf{r}$ being the position vector ($\mathbf{r}=x\hat{\mathbf{u}}_{x}+y\hat{\mathbf{u}}_{y}+z\hat{\mathbf{u}}_{z}$,
$\hat{\mathbf{u}}_{i}$ being the unit vector along the $i$-th ($i=\left\{ x,y,z\right\} $)
direction), univocally described by a set of $N$ real-valued parameters,
$\underline{p}=\left\{ p_{n}\in\mathbb{R};\, n=1,...,N\right\} $,
the $n$-th ($n=1,...,N$) one being not exactly known, since affected
by tolerance, thus belonging to a bounded/finite interval $\left[p_{n}\right]$
(i.e., $p_{n}\in\left[p_{n}\right]$) \cite{Moore 1966} defined as
follows\begin{equation}
\left[p_{n}\right]=\left[\inf\left(p_{n}\right);\sup\left(p_{n}\right)\right]\label{eq:_parameter-interval.LBE-input}\end{equation}
where $\inf\left(p_{n}\right)=p_{n}-\inf\left(\delta_{n}\right)$
and $\sup\left(p_{n}\right)=p_{n}+\sup\left(\delta_{n}\right)$ are
the left/lower and the right/upper endpoints, respectively, $\inf\left(\delta_{n}\right)$
and $\sup\left(\delta_{n}\right)$ being the uncertainties with respect
to the nominal (i.e., tolerance/error-free) parameter value $p_{n}$.
Furthermore, let the \emph{TF} $\mathcal{F}$ between the parameter
vector $\underline{p}$ and the response of the \emph{EM} device,
$\Phi\left(\theta,\phi;\underline{p}\right)$ (Fig. 1), $\left(\theta,\phi\right)$
being a generic set of angular coordinates, namely\begin{equation}
\Phi\left(\theta,\phi;\underline{p}\right)=\mathcal{F}\left(\underline{p}\right),\label{eq:_response.LBE-output}\end{equation}
be not explicitly/analytically known, even though a set of $S$ input-output
(I/O) examples, \{($\underline{p}^{\left(s\right)}$; $\Phi^{\left(s\right)}$);
$s=1,...,S$\} {[}$\Phi^{\left(s\right)}\triangleq\Phi\left(\theta,\phi;\underline{p}^{\left(s\right)}\right)${]},
are available (i.e., there is the so-called {}``\emph{implicit}''
knowledge of the \emph{TF}).

\noindent We are then interested in evaluating the response of the
\emph{EM} device (e.g., the far-field power pattern in an antenna
system), $\Phi\left(\theta,\phi;\underline{p}\right)$, when its descriptors,
$\underline{p}$, are affected by tolerances. Mathematically, the
sensitivity analysis problem at hand can be stated as follows:

\begin{quote}
\emph{Generalized IA-LBE Sensitivity Analysis} (\emph{GILSA}) \emph{Problem}
- Given a set of $N$ real-valued descriptors, $\underline{p}=\left\{ p_{n};\, n=1,...,N\right\} $,
of the representative function of a generic \emph{EM} device, $\mathbf{J}\left(\mathbf{r}\,;\underline{p}\right)$,
which are characterized by uncertain, but bounded, tolerances such
that their actual values lay within known intervals, $\left[\underline{p}\right]$
$=$ \{$\left[p_{n}\right]$; $n=1,...,N$\}, (i.e., $\underline{p}\in\left[\underline{p}\right]$)
and a set of $S$ known I/O pairs, \{($\underline{p}^{\left(s\right)}$;
$\Phi^{\left(s\right)}$); $s=1,...,S$\}, $\Phi^{\left(s\right)}=\mathcal{F}\left(\underline{p}^{\left(s\right)}\right)$
being the response of the \emph{EM} device with unknown \emph{TF}
$\mathcal{F}$, derive an approximated explicit expression $\widetilde{\mathcal{F}}$
of $\mathcal{F}$ (Fig. 2) and extend it to interval by means of the
\emph{IA} as\begin{equation}
\widetilde{\mathcal{F}}\left(\left[\underline{p}\right]\right)=\widetilde{\Phi}\left(\theta,\phi;\left[\underline{p}\right]\right)\label{eq:_approx-response.IA}\end{equation}
to yield the analytic expressions of the corresponding lower, $\inf\left\{ \widetilde{\Phi}\left(\theta,\phi\right)\right\} $,
and upper, $\sup\left\{ \widetilde{\Phi}\left(\theta,\phi\right)\right\} $,
bounds ($\widetilde{\Phi}\left(\theta,\phi;\left[\underline{p}\right]\right)=\left[\inf\left\{ \widetilde{\Phi}\left(\theta,\phi\right)\right\} ;\sup\left\{ \widetilde{\Phi}\left(\theta,\phi\right)\right\} \right]$)
that include the actual response of the \emph{EM} device {[}$\mathcal{F}\left(\underline{p}'\right)\subset\widetilde{\Phi}\left(\theta,\phi;\left[\underline{p}\right]\right)${]}
for any admissible combination of the parameter values, $\underline{p}'$,
laying within the tolerance intervals (i.e., $\underline{p}'\in\left[\underline{p}\right]$).
\end{quote}

\section{\emph{IA-LBE} Sensitivity Analysis Method\label{sec:IA-LBE-Sensitivity-Analysis}}

In order to solve the \emph{GILSA} problem, the proposed sensitivity
analysis method (\emph{IA-LBE}) is based on a suitable integration
of \emph{LBE} and \emph{IA} techniques. First, a \emph{LBE} approach
is exploited to derive a \emph{SM} of the \emph{TF, $\widetilde{\mathcal{F}}\left(\underline{p}\right)$}
{[}$\widetilde{\mathcal{F}}\left(\underline{p}\right)\cong\mathcal{F}\left(\underline{p}\right)${]},
starting from the knowledge of a training set of $S$ I/O pairs, \{($\underline{p}^{\left(s\right)}$;
$\Phi^{\left(s\right)}$); $s=1,...,S$\}. Then, the \emph{IA} is
applied to extend to intervals the analytic expression of the \emph{SM,
$\widetilde{\mathcal{F}}\left(\underline{p}\right)$,} {[}$\widetilde{\mathcal{F}}\left(\underline{p}\right)\to\widetilde{\mathcal{F}}\left(\left[\underline{p}\right]\right)$
(\ref{eq:_approx-response.IA}){]} by deriving the closed-form expressions
of the corresponding endpoints, $\inf\left\{ \widetilde{\Phi}\left(\theta,\phi\right)\right\} $
and $\sup\left\{ \widetilde{\Phi}\left(\theta,\phi\right)\right\} $.

\subsection{\noindent \emph{TF LBE}-Based Explicit Model\label{sub:TF-LBE-Based-Explicit}}

\noindent To derive the analytic expression of the \emph{SM} $\widetilde{\mathcal{F}}\left(\cdot\right)$,
the Ordinary Kriging (\emph{OK}) technique is used as \emph{LBE} method
because of its suitability when dealing with high-fidelity I/O pairs,
\{($\underline{p}^{\left(s\right)}$; $\Phi^{\left(s\right)}$); $s=1,...,S$\},
as those generated with full-wave simulations \cite{Salucci 2022},
since the \emph{OK} interpolates the training samples to build the
\emph{SM}. More specifically, the \emph{SM} $\widetilde{\mathcal{F}}\left(\underline{p}\right)$
is given by \cite{Massa 2018}\cite{Massa 2022}-\cite{Jones 1998}\begin{equation}
\widetilde{\mathcal{F}}\left(\underline{p}\right)=\gamma+\sum_{a=1}^{S}\sum_{b=1}^{S}c\left(\underline{p},\underline{p}^{\left(a\right)}\right)w_{ab}\left(\Phi^{\left(b\right)}-\gamma\right)\label{eq:_OK-prediction}\end{equation}
where $w_{ab}$ is the $\left(a,b\right)$-th ($a=1,...,S$; $b=1,...,S$)
weighting coefficient as well as an entry of the matrix $\underline{\underline{W}}\in\mathbb{R}^{S\times S}$
($\underline{\underline{W}}\triangleq\left\{ w_{ab};\, a=1,...,S;\, b=1,...,S\right\} $),
which is the inverse of the known training correlation matrix $\underline{\underline{C}}\in\mathbb{R}^{S\times S}$($\underline{\underline{C}}\triangleq\left\{ c_{ab}=c\left(\underline{p}^{\left(a\right)},\underline{p}^{\left(b\right)}\right);\, a=1,...,S;\, b=1,...,S\right\} $)
(i.e., $\underline{\underline{W}}=\underline{\underline{C}}^{-1}$),
and $\gamma$ is the known regression term equal to\begin{equation}
\gamma=\frac{\sum_{a=1}^{S}\sum_{b=1}^{S}w_{ab}\Phi^{\left(b\right)}}{\sum_{a=1}^{S}\sum_{b=1}^{S}w_{ab}}.\label{eq:_OK-regression-term}\end{equation}
Moreover, $c\left(\underline{p},\underline{p}^{\left(a\right)}\right)$
is the $a$-th ($a=1,...,S$) correlation coefficient between the
unknown input $\underline{p}$ and the $a$-th ($a=1,...,S$) training
sample, $\underline{p}^{\left(a\right)}$, which is defined as follows

\noindent \begin{equation}
c\left(\underline{p},\underline{p}^{\left(a\right)}\right)=\exp\left\{ -\sum_{n=1}^{N}\beta_{n}\left|\underline{p}_{n}^{\left(a\right)}-p_{n}\right|^{\alpha_{n}}\right\} \label{eq:_correlation-training-input}\end{equation}
where $\alpha_{n}$ and $\beta_{n}$ ($n=1,...,N$) are \emph{OK}
control hyper-parameters.

\subsection{\noindent \emph{IA}-Based Interval Bounds Derivation\label{sub:IA-Based-Interval-Bounds}}

\noindent Since in (\ref{eq:_OK-prediction}), the only term affected
by tolerance is $c\left(\underline{p},\underline{p}^{\left(a\right)}\right)$
being a function of the input vector $\underline{p}$, while all other
terms are known \emph{crisp} values that are computed from the known
$S$-size training set, \{($\underline{p}^{\left(s\right)}$; $\Phi^{\left(s\right)}$);
$s=1,...,S$\}, the interval extension%
\footnote{\noindent The function $\mathcal{G}\left(\left[\chi\right]\right)$
is called interval extension of $\mathcal{G}\left(\chi\right)$ if
for degenerate interval arguments, namely when $\left[\chi\right]=\left[\inf\left(\chi\right);\,\sup\left(\chi\right)\right]=\chi$
(i.e., the left and right endpoints coincide, $\inf\left(\chi\right)=\sup\left(\chi\right)=\chi$),
$\mathcal{G}\left(\left[\chi\right]\right)=\mathcal{G}\left(\chi\right)$,
$\forall\chi\in\left[\chi\right]$ \cite{Moore 1966}.%
} of (\ref{eq:_OK-prediction}) is yielded by substituting the parameter
intervals, \{$\left[p_{n}\right]$; $n=1,...,N$\}, in (\ref{eq:_correlation-training-input})\begin{equation}
\widetilde{\mathcal{F}}\left(\left[\underline{p}\right]\right)=\gamma+\sum_{a=1}^{S}\sum_{b=1}^{S}\exp\left\{ -\sum_{n=1}^{N}\beta_{n}\left|\underline{p}_{n}^{\left(a\right)}-\left[p_{n}\right]\right|^{\alpha_{n}}\right\} w_{ab}\left(\Phi^{\left(b\right)}-\gamma\right).\label{eq:_OK-interval-prediction}\end{equation}
Since the $a$-th ($a=1,...,S$) interval exponent of (\ref{eq:_OK-interval-prediction}),\begin{equation}
\left[\Theta^{\left(a\right)}\right]=\sum_{n=1}^{N}\beta_{n}\left|\underline{p}_{n}^{\left(a\right)}-\left[p_{n}\right]\right|^{\alpha_{n}},\label{eq:_interval-exponent}\end{equation}
is a linear combination of $N$ terms, then its endpoints, $\inf\left\{ \Theta^{\left(a\right)}\right\} $
and $\sup\left\{ \Theta^{\left(a\right)}\right\} $ ($\left[\Theta_{\left(a\right)}\right]=\left[\inf\left\{ \Theta^{\left(a\right)}\right\} ;\sup\left\{ \Theta^{\left(a\right)}\right\} \right]$)
are derived by applying the \emph{IA} sum rule \cite{Moore 1966}
(\emph{Appendix I}) so it turns out that\begin{equation}
\frac{\inf}{\sup}\left\{ \Theta^{\left(a\right)}\right\} =\sum_{n=1}^{N}\beta_{n}\frac{\inf}{\sup}\left\{ D_{n}^{\left(a\right)}\right\} \label{eq:_interval-exponent-inf}\end{equation}
where $\inf\left\{ D_{n}^{\left(a\right)}\right\} $ and $\sup\left\{ D_{n}^{\left(a\right)}\right\} $
are the endpoints of the interval $\left[D_{n}^{\left(a\right)}\right]$
($\left[D_{n}^{\left(a\right)}\right]\triangleq\left|p_{n}^{\left(a\right)}-\left[p_{n}\right]\right|^{\alpha_{n}}$).

\noindent These latter are determined by applying the \emph{IA} rule
for the power of the absolute value of an interval \cite{Moore 1966}
(\emph{Appendix II}) so that

\noindent \begin{equation}
\frac{\inf}{\sup}\left\{ D_{n}^{\left(a\right)}\right\} =\left\{ \begin{array}{lll}
\left\{ p_{n}^{\left(a\right)}-\frac{\sup}{\inf}\left(p_{n}\right)\right\} ^{\alpha_{n}} & if & \left\{ \sup\left(p_{n}\right)\right\} <p_{n}^{\left(a\right)}\\
\left\{ p_{n}^{\left(a\right)}-\frac{\inf}{\sup}\left(p_{n}\right)\right\} ^{\alpha_{n}} & if & \left\{ \inf\left(p_{n}\right)\right\} >p_{n}^{\left(a\right)}\\
0 & if & \left\{ \inf\left(p_{n}\right)<p_{n}^{\left(a\right)}\right\} \,\,\&\,\,\left\{ \sup\left(p_{n}\right)>p_{n}^{\left(a\right)}\right\} \end{array}\right.\label{eq:_inf_power_distance}\end{equation}
where $\max\left(D_{n}\right)=\max\left(\left\{ p_{n}^{\left(a\right)}-\inf\left(p_{n}\right)\right\} ^{\alpha_{n}},\,\left\{ p_{n}^{\left(a\right)}-\sup\left(p_{n}\right)\right\} ^{\alpha_{n}}\right)$,
while $\&$ stands for the {}``and'' Boolean operator.

\noindent Finally, since the exponential function is monotonic%
\footnote{\noindent Any monotonic function $\mathcal{G}\left(\chi\right)$ can
be directly extended to its interval counterpart, $\mathcal{G}\left(\left[\chi\right]\right)$,
as follows: $\mathcal{G}\left(\left[\chi\right]\right)=\left[\mathcal{G}\left(\inf\left(\chi\right)\right);\mathcal{G}\left(\sup\left(\chi\right)\right)\right]$
or $\mathcal{G}\left(\left[\chi\right]\right)=\left[\mathcal{G}\left(\sup\left(\chi\right)\right);\mathcal{G}\left(\inf\left(\chi\right)\right)\right]$
if the function $\mathcal{G}\left(\chi\right)$ increases or decreases
with respect to its argument, $\chi$, respectively \cite{Moore 1966}.%
}, the analytic expressions of the lower and the upper bounds of the
interval \emph{SM} (\ref{eq:_OK-interval-prediction}), $\widetilde{\mathcal{F}}\left(\left[\underline{p}\right]\right)$
(\ref{eq:_approx-response.IA}), are given by\begin{equation}
\frac{\inf}{\sup}\left\{ \widetilde{\Phi}\left(\theta,\phi\right)\right\} =\gamma+\sum_{a=1}^{S}\Lambda_{a}\exp\left(-\sum_{n=1}^{N}\beta_{n}\frac{\sup}{\inf}\left\{ D_{n}^{\left(a\right)}\right\} \right)\label{eq:_OK-interval-prediction-inf}\end{equation}
where $\Lambda_{a}=\sum_{b=1}^{S}w_{ab}\left(\Phi^{\left(b\right)}-\gamma\right)$
is a crisp term only function of the $S$ training examples, \{($\underline{p}^{\left(s\right)}$;
$\Phi^{\left(s\right)}$); $s=1,...,S$\}.

\section{Numerical Results\label{sec:Numerical-Results}}

This section is devoted to illustrate the features of the proposed
sensitivity analysis approach as well as to assess its performance.
As a representative example of \emph{EM} devices, antenna systems
are considered in the numerical validation and different reference
examples are dealt with. More specifically, after a calibration of
its control parameters on an analytic benchmark function, the \emph{IA-LBE}
is first validated in selected test cases where the explicit form
of the \emph{TF} is known to verify the closeness of the arising bounds
to the ideal ones of the exact \emph{IA} (\emph{IA}-\emph{E}). Successively,
the proposed approach is applied to a realistic antenna element for
which the I/O relationship is not explicitly known, but only a set
of $S$ I/O examples is available.

\noindent As for the validation of the \emph{IA-LBE} method, the first
numerical example considers an analytic benchmark function\begin{equation}
\Phi\left(\theta;\underline{p}\right)=\sum_{n=1}^{N}p_{n}\theta^{n}\label{eq:_EX-1.IO-function}\end{equation}
where the $n$-th ($n=1,...N$) control parameter, $p_{n}$, is affected
by an uncertainty up to $\frac{\frac{\inf}{\sup}\left(\delta_{n}\right)}{p_{n}}=\delta=20\%$
with respect to the nominal unitary value ($p_{n}=1.0$) so that $p_{n}\in\left[p_{n}\right]$,
$\inf\left(p_{n}\right)=p_{n}\times\left(1-\delta\right)$ and $\sup\left(p_{n}\right)=p_{n}\times\left(1+\delta\right)$
being the endpoints of $\left[p_{n}\right]$. The values of the input
parameters, \{$\underline{p}^{\left(s\right)}$; $s=1,...,S$\}, of
the $S$ I/O training pairs for defining the \emph{SM} of the \emph{TF}
(\ref{eq:_OK-prediction}) have been selected by sampling each $n$-th
($n=1,...,N$) interval $\left[p_{n}\right]$ with the Latin hypercube
sampling (\emph{LHS}) strategy \cite{Liu 2017}. Moreover, since the
output function (\ref{eq:_EX-1.IO-function}) varies continuously
with $\theta$, $K=201$ uniformly-distributed $\theta$-samples have
been taken within the domain $\Omega\triangleq\left\{ \theta:\left|\theta\right|\leq1\right\} $,
namely \{$\Phi\left(\theta_{k};\underline{p}\right)$; $k=1,...,K$\}
where $\theta_{k}=-1+2\left(\frac{k-1}{K-1}\right)$. Next, a \emph{SM}
has been trained according to the procedure detailed in Sect. \ref{sub:TF-LBE-Based-Explicit}
for every $k$-th ($k=1,...,K$) angular sample, $\theta_{k}$, with
the corresponding training set of $S$ I/O pairs, \{$\left(\underline{p}^{\left(s\right)};\Phi_{k}^{\left(s\right)}\right)$;
$s=1,...,S$\} {[}$\Phi_{k}^{\left(s\right)}\triangleq\Phi\left(\theta_{k};\underline{p}^{\left(s\right)}\right)${]}
by setting the \emph{OK} hyper-parameter $\alpha_{n}$ ($n=1,...,N$)
to $\alpha_{n}=2$, while optimizing the other ones, \{$\beta_{n}$;
$n=1,...,N$\}, by means of a local search optimization \cite{Kowalik 1968}.
Owing to the stochastic nature of the \emph{LHS}, $L=10^{4}$ different
samplings of $\left[\underline{p}\right]$, \{$\underline{p}_{l}^{\left(s\right)}$;
$l=1,...,L$\}, have been carried out and $L$ \emph{SM}s, \{$\widetilde{\mathcal{F}}_{l}\left(\left[\underline{p}\right]\right)$;
$l=1,...,L$\} {[}$\widetilde{\mathcal{F}}_{l}\left(\left[\underline{p}\right]\right)\triangleq\widetilde{\Phi}_{l}\left(\theta;\left[\underline{p}\right]\right)${]},
have been built, but the one with minimum \emph{tolerance index} $\Delta$
\cite{Anselmi 2013}, that is\begin{equation}
\widetilde{\Phi}\left(\theta;\left[\underline{p}\right]\right)\equiv\begin{array}{r}
\arg\min_{l=1,...,L}\Delta\left\{ \widetilde{\Phi}_{l}\left(\theta;\left[\underline{p}\right]\right)\right\} \end{array},\label{eq:_min-tolerance-index}\end{equation}
has been kept for the successive analysis. In (\ref{eq:_min-tolerance-index}),
$\Delta\left\{ \widetilde{\Phi}_{l}\left(\theta;\left[\underline{p}\right]\right)\right\} =\frac{\int_{\Omega}\left|\sup\left\{ \widetilde{\Phi}_{l}\left(\theta\right)\right\} -\inf\left\{ \widetilde{\Phi}_{l}\left(\theta\right)\right\} \right|d\theta}{\int_{\Omega}\left|\Phi\left(\theta\right)\right|d\theta}$.
Such a choice (\ref{eq:_min-tolerance-index}) is motivated by the
fact that the \emph{SM} with the minimum $\Delta$ value has the closest
interval bounds, which is the most critical case when checking the
inclusion property%
\footnote{\noindent The \emph{inclusion property} of \emph{IA} states that,
whatever interval function evaluated over an interval argument {[}i.e.,
$\mathcal{G}\left(\left[\chi\right]\right)${]}, it includes all possible
values computed at sample points of the interval argument, namely
$\mathcal{G}\left(\chi\right)\subseteq\mathcal{G}\left(\left[\chi\right]\right)$
for $\forall\chi\in\left[\chi\right]$.%
} \cite{Moore 1966}\cite{Hansen 2004} (i.e., if the \emph{IA} bounds
include all \emph{MC} realizations).

\noindent Figure 3 compares the lower/upper bounds of the \emph{IA-LBE}
method (\ref{eq:_OK-interval-prediction-inf}) with $M=10^{6}$ \emph{MC}
realizations of the prediction process {[}$\underline{p}\to\Phi\left(\theta;\underline{p}\right)=\mathcal{F}\left(\underline{p}\right)${]}
when $N=1$. The \emph{IA}-\emph{E} bounds, which have been computed
by directly applying the arithmetic of intervals to (\ref{eq:_EX-1.IO-function}),
are reported, as well. Moreover, $S$ has been varied within the range
$1\leq\frac{S}{N}\leq8$ to calibrate the size of the training set.
It turns out that the \emph{IA-LBE} bounds are narrower than both
the \emph{E-IA} bounds and the \emph{MC} curves until $S=5\times N$,
while the inclusion property is fulfilled when $S\geq6\times N$.
To quantitatively highlight such an outcome, let us introduce the
\emph{interval inclusion metric} $\Psi$ defined as\begin{equation}
\begin{array}{r}
\Psi\triangleq\left\{ \begin{array}{ccc}
\Psi_{ext} & if & \Psi_{int}=0\\
-\left(\Psi_{int}+\Psi_{pen}\right) & if & \Psi_{int}\neq0\end{array}\right.\end{array}\label{eq:_interval-inclusion-metric}\end{equation}
where the terms $\Psi_{int}$,\begin{equation}
\begin{array}{ccc}
\Psi_{int} & \triangleq & \frac{\int_{\Omega}\left(\inf\left\{ \Phi_{IA}\left(\theta\right)\right\} -\inf\left\{ \Phi_{MC}\left(\theta\right)\right\} \right)\times\mathcal{H}\left(\inf\left\{ \Phi_{IA}\left(\theta\right)\right\} -\inf\left\{ \Phi_{MC}\left(\theta\right)\right\} \right)d\theta}{\int_{\Omega}\left(\sup\left\{ \Phi_{MC}\left(\theta\right)\right\} -\inf\left\{ \Phi_{MC}\left(\theta\right)\right\} \right)d\theta}+\\
 &  & \frac{\int_{\Omega}\left(\sup\left\{ \Phi_{MC}\left(\theta\right)\right\} -\sup\left\{ \Phi_{IA}\left(\theta\right)\right\} \right)\times\mathcal{H}\left(\sup\left\{ \Phi_{MC}\left(\theta\right)\right\} -\sup\left\{ \Phi_{IA}\left(\theta\right)\right\} \right)d\theta}{\int_{\Omega}\left(\sup\left\{ \Phi_{MC}\left(\theta\right)\right\} -\inf\left\{ \Phi_{MC}\left(\theta\right)\right\} \right)d\theta},\end{array}\label{eq:_inclusion-metric.interna}\end{equation}
and $\Psi_{ext}$,\begin{equation}
\begin{array}{r}
\begin{array}{ccc}
\Psi_{ext} & \triangleq & \frac{\int_{\Omega}\left(\inf\left\{ \Phi_{MC}\left(\theta\right)\right\} -\inf\left\{ \Phi_{IA}\left(\theta\right)\right\} \right)\times\mathcal{H}\left(\inf\left\{ \Phi_{MC}\left(\theta\right)\right\} -\inf\left\{ \Phi_{IA}\left(\theta\right)\right\} \right)d\theta}{\int_{\Omega}\left(\sup\left\{ \Phi_{MC}\left(\theta\right)\right\} -\inf\left\{ \Phi_{MC}\left(\theta\right)\right\} \right)d\theta}+\\
 &  & \frac{\int_{\Omega}\left(\sup\left\{ \Phi_{IA}\left(\theta\right)\right\} -\sup\left\{ \Phi_{MC}\left(\theta\right)\right\} \right)\times\mathcal{H}\left(\sup\left\{ \Phi_{IA}\left(\theta\right)\right\} -\sup\left\{ \Phi_{MC}\left(\theta\right)\right\} \right)d\theta}{\int_{\Omega}\left(\sup\left\{ \Phi_{MC}\left(\theta\right)\right\} -\inf\left\{ \Phi_{MC}\left(\theta\right)\right\} \right)d\theta}\end{array}\end{array},\label{eq:_inclusion-metric.esterna}\end{equation}
quantify how much the \emph{IA} bounds {[}i.e., either \emph{IA-LBE}
bounds, $\Phi_{IA}\leftarrow\widetilde{\Phi}_{IA-LBE}$, or \emph{IA-E}
bounds, $\Phi_{IA}\leftarrow\Phi_{IA-E}${]} are internal/under-estimated
or external/over-estimated with respect to the \emph{MC} band whose
{}``endpoints'', $\inf\left\{ \Phi_{MC}\left(\theta\right)\right\} $
and $\sup\left\{ \Phi_{MC}\left(\theta\right)\right\} $, are given
by\begin{equation}
\frac{\inf}{\sup}\left\{ \Phi_{MC}\left(\theta\right)\right\} =\frac{\min}{\max}_{m=1,...M}\left\{ \Phi_{MC}^{\left(m\right)}\left(\theta\right)\right\} ,\label{eq:}\end{equation}
$\Phi_{MC}^{\left(m\right)}\left(\theta\right)$ being the $m$-th
($m=1,...,M$) \emph{MC} realization, while $\mathcal{H}\left(\cdot\right)$
is the Heaviside function. Moreover,\begin{equation}
\begin{array}{r}
\begin{array}{ccc}
\Psi_{pen} & \triangleq & \frac{\int_{\Omega}\left(\inf\left\{ \Phi_{IA}\left(\theta\right)\right\} -\Phi\left(\theta\right)\right)\times\mathcal{H}\left(\inf\left\{ \Phi_{IA}\left(\theta\right)\right\} -\Phi\left(\theta\right)\right)d\theta}{\int_{\Omega}\left(\sup\left\{ \Phi_{MC}\left(\theta\right)\right\} -\inf\left\{ \Phi_{MC}\left(\theta\right)\right\} \right)d\theta}+\\
 &  & \frac{\int_{\Omega}\left(\Phi\left(\theta\right)-\sup\left\{ \Phi_{IA}\left(\theta\right)\right\} \right)\times\mathcal{H}\left(\Phi\left(\theta\right)-\sup\left\{ \Phi_{IA}\left(\theta\right)\right\} \right)d\theta}{\int_{\Omega}\left(\sup\left\{ \Phi_{MC}\left(\theta\right)\right\} -\inf\left\{ \Phi_{MC}\left(\theta\right)\right\} \right)d\theta}\end{array}\end{array}\label{eq:_inclusion-metirc.wrt-nominal}\end{equation}
is a penalty term equal to the area between the \emph{IA} bounds and
the nominal (tolerance-free) function $\Phi$ when this latter is
not included in the \emph{IA} bounds.

\noindent Figure 4 shows the behaviour of $\Psi$ versus the (normalized)
number of training samples, $\frac{S}{N}$, for both the \emph{IA-LBE}
and the \emph{IA-E}. As expected, the \emph{IA-E} bounds always include
the \emph{MC} predictions by virtue of the \emph{inclusion property}
of \emph{IA}, the value of the \emph{interval inclusion metric} (\ref{eq:_interval-inclusion-metric})
being always positive and equal to $\Psi_{IA-E}=1.01\times10^{-2}$
(Fig. 4). Differently, the value of $\Psi_{IA-LBE}$ is \emph{}negative
when $\frac{S}{N}<6$ (e.g., $\left.\Psi_{IA-LBE}\right\rfloor _{\frac{S}{N}=1}=-1.51$
and $\left.\Psi_{IA-LBE}\right\rfloor _{\frac{S}{N}=5}=-3.10\times10^{-1}$
- Tab. I) because of the under-estimation of the extension of the
\emph{MC} band, while it becomes positive over $\frac{S}{N}=6$ (e.g.,
$\left.\Psi_{IA-LBE}\right\rfloor _{\frac{S}{N}=6}=8.35\times10^{-2}$
- Tab. I) with greater and greater values as the training set size
increases (i.e., $\frac{S}{N}$). Therefore, the number of training
samples has been set to $S=S_{th}$ ($S_{th}=6\times N$) hereinafter.

\noindent As for the computational costs, the \emph{IA-LBE} approach
took $\tau_{IA-LBE}=0.1$ {[}sec{]}, including the time for the generation
of the $S=S_{th}$ I/O samples, while the Monte Carlo simulations
required $\tau_{MC}=13.5$ {[}sec{]} %
\footnote{\noindent On a standard laptop equipped with an Intel-i5 CPU @ 1.60
GHz and 8 GB of RAM.%
}. It means a time-saving of $\Delta\tau_{MC}^{IA-LBE}\approx99.2$
\%. For completeness, $\tau_{IA-E}=7\times10^{-3}$ {[}sec{]}.

\noindent The second experiment of the first test case deals with
the dependence of $\Psi$ on the tolerance error, which is varied
within the range $5$ \% $\le$ $\delta$ $\le$ $50$ \%. The results
in Fig. 5 prove the reliability of the setup $S=S_{th}$ ($S_{th}=6\times N$)
since $\Psi_{IA-LBE}$ is always positive and constant versus the
tolerance value, $\delta$, as well as close to the asymptotic bound
$\Psi_{IA-E}$. For illustrative purposes, Figure 6 shows the plots
of the \emph{IA-LBE} bounds, the \emph{IA-E} bounds, and the $M$
\emph{MC} predictions when $\delta=5$ \% {[}Fig. 6(\emph{a}){]},
$\delta=10$ \% {[}Fig. 6(\emph{b}){]}, $\delta=30$ \% {[}Fig. 6(\emph{c}){]},
$\delta=40$ \% {[}Fig. 6(\emph{d}){]}, and $\delta=50$ \% {[}Fig.
6(\emph{e}){]}.

\noindent In the last experiment of this numerical example, the number
of parameters affected by uncertainty has been increased up to $N=10$,
while fixing the tolerance value to $\delta=20$ \%. The behaviour
of the \emph{interval inclusion metric}, $\Psi$, versus $N$ as well
as the pictorial comparison between the \emph{IA-LBE} bounds, the
\emph{IA-E} bounds, and the $M$ \emph{MC} realizations for different
values of $N$ are reported in Fig. 7 and Fig. 8, respectively. While
both figures confirm the outcomes drawn from Figs. 5-6 on the fulfillment
of the inclusion property as well as the validity of the choice on
the size of the training set (i.e., $S=S_{th}$, $S_{th}=6\times N$),
Figure 7 points out a steady growing trend of the $\Psi$ value with
respect to $N$. Such a behavior, which is more evident in the \emph{IA-LBE}
result, is due to the intrinsic over-estimation of the bounds caused
by the \emph{IA} {}``wrapping'' effect \cite{Tenuti 2017} when
more and more intervals are summed as it happens in (\ref{eq:_EX-1.IO-function})
when $N$ increases.

\noindent The second example is concerned with a linear array of $N=10$
isotropic antennas whose excitation amplitudes, \{$p_{n}$; $n=1,...,N$\},
deviate from the nominal values. As in the previous test case, the
\emph{TF} of such an \emph{EM} device is explicitly known and it is
the radiated far-field power pattern\begin{equation}
\Phi\left(\theta;\underline{p}\right)=\left|\sum_{n=1}^{N}p_{n}e^{j\frac{2\pi}{\lambda}d\left(n-1\right)\sin\theta}\right|^{2}\label{eq:_EX-2.IO-function}\end{equation}
where $\lambda$ is the wavelength, $d=\frac{\lambda}{2}$ is the
inter-element spacing between two adjacent elementary radiators located
along the (array) $x$-axis, and $\theta\in\left[-\frac{\pi}{2};\frac{\pi}{2}\right]$
is the angular direction measured from broadside \cite{Mailloux 2018}.
The nominal values of \{$p_{n}$; $n=1,...N$\} have been set to the
Dolph-Chebyshev \cite{Dolph 1046} ones to afford a power pattern
with \emph{SLL} equal to $SLL=-20$ {[}dB{]}.

\noindent In the first numerical experiment, tolerances in the range
$1$ \% $\le$$\delta$ $\le$ $10$ \% have been considered as sketched
in Fig. 9 where the nominal values are displayed, as well. Moreover,
the array power pattern (\ref{eq:_EX-2.IO-function}) has been sampled
into $K=380$ points (i.e., $20$ times the Nyquist rate equal to
$2N-1$ \cite{Liu 2008}) uniformly chosen in the angular domain $\Omega$
($\Omega\triangleq\left\{ \theta:\left|\theta\right|\leq\frac{\pi}{2}\right\} $).
The \emph{SM}, $\widetilde{\mathcal{F}}\left(\underline{p}\right)$,
of (\ref{eq:_EX-2.IO-function}) and the \emph{IA-LBE} bounds have
been derived by using $S=60$ (i.e., $S=S_{th}$, $S_{th}=6\times N$,
$N$ being equal to $N=10$) \emph{LHS} training samples and keeping
the same setting of the \emph{OK} hyper-parameters of the first example.
Figure 10 gives the plots of $\Psi$ versus the tolerance level. Whatever
the value of $\delta$, the \emph{IA-LBE} bounds turn out to be always
inclusive of the \emph{MC} band (i.e., $\Psi_{IA-LBE}>0$) and close
to the \emph{IA-E} bounds (Tab. II). Moreover, the value of the interval
inclusion metric is almost constant within the whole range of $\delta$
variations with a decreasing behaviour as $\delta$ increases. For
illustrative purposes, some representative plots of the \emph{IA-LBE}
bounds together with the \emph{IA-E} ones%
\footnote{\noindent The analytic formulas of the \emph{IA-E} interval power
pattern bounds are available in \cite{Anselmi 2013}.%
} and the \emph{MC} simulations are reported in Fig. 11, while the
intervals of the main pattern features \cite{Anselmi 2013} (i.e.,
the \emph{SLL}, the half-power beamwidth (\emph{BW}), the power peak
$\Phi_{max}$ ($\Phi_{max}\triangleq\max_{\theta}\left\{ \Phi\left(\theta\right)\right\} $),
and the \emph{tolerance index} $\Delta$ \cite{Anselmi 2013}) are
reported in Tab. II to give to the interested readers some more information
on the method outcomes. As for these latter, it is worth noticing
that the nominal values of every power pattern feature lie within
the corresponding \emph{IA-LBE} bounds (Tab. II).

\noindent Similar outcomes arise when dealing with larger arrays (i.e.,
$N>10$). For instance, Figure 12 is concerned with an array with
$N=20$ {[}Figs. 12(\emph{a})-12(\emph{c}){]} and $N=50$ {[}Figs.
12(\emph{b})-12(\emph{d}){]} $d=\frac{\lambda}{2}$-space elements
where the nominal excitations, which generate a Dolph-Chebyshev with
$SLL=-20$ {[}dB{]}, have been corrupted with a $\delta=10$ \% tolerance.
Analogously to the example of the benchmark function (\ref{eq:_EX-1.IO-function}),
but now for a different \emph{TF} (\ref{eq:_EX-2.IO-function}), the
value of $\Psi$ is monotonically increasing with the number of parameters
affected by uncertainty, $N$ (here the number of array elements),
(Fig. 13) for both \emph{IA}-based methods (e.g., $\left.\Psi_{IA-LBE}\right\rfloor _{N=20}=1.89$
vs. $\left.\Psi_{IA-E}\right\rfloor _{N=20}=1.21$ and $\left.\Psi_{IA-LBE}\right\rfloor _{N=50}=3.10$
vs. $\left.\Psi_{IA-E}\right\rfloor _{N=50}=2.22$ - Tab. III).

\noindent The last example is devoted to assess the proposed method
in its entirety and complete operational capability, that is when
the explicit form of the \emph{TF} $\mathcal{F}\left(\underline{p}\right)$
is unknown. Towards this end, the tolerance analysis of the single-polarization
aperture-coupled stacked square patch \emph{}antenna \cite{Targonski 1998}
in Fig. 14 has been carried out. More in detail, the \emph{}antenna
consists of two stacked microstrip patch radiators printed on top
of two substrates and is fed by a microstrip line coupled through
a rectangular slot to resonate at $f=3.5$ {[}GHz{]}. The slot is
etched in a ground-plane that separates the patches and the feeding
line, while the dielectric material of the antenna substrates is Rogers
{}``RT/Duroid 3003'' \cite{Rogers 2022} with relative permittivity
and loss tangent equal to $\varepsilon_{r}=3.0$ and $\tan\delta=0.0016$,
respectively. The antenna descriptors and their nominal values are
indicated in Fig. 14 and reported in Tab. IV, respectively.

\noindent In the first experiment, the sensitivity of the realized
gain (\emph{RG}) pattern from a deviation of $\delta=3\%$ (Tab. V)
by the nominal value of the thickness of the lowest substrate layer
$h_{0}$ has been analyzed ($n=N=1$ and $p_{n}=h_{0}$). Accordingly,
a set of $S=6$ ($S=S_{th}$, $S_{th}=6\times N$) co-polar \emph{RG}
patterns has been simulated with a commercial full-wave \emph{EM}
software \cite{HFSS 2021} by discretizing the pattern along the principal
planes $\phi=0$ {[}deg{]} and $\phi=90$ {[}deg{]} into $K=361$
directions within the domain $\Omega\triangleq\left\{ \theta:\left|\theta\right|\leq180\right\} $
{[}deg{]}, the direction $\theta=0$ {[}deg{]} being that of the $z$-axis.
Also in this case, the \emph{LHS}-based sampling of the input interval
vector $\left[\underline{p}\right]$ has been repeated $L=10^{4}$
times and the \emph{SM} has been selected as in (\ref{eq:_min-tolerance-index}).

\noindent Figure 15 shows in the $\phi=0$ {[}deg{]} plane (Fig. 15
- \emph{first row}) and the $\phi=90$ {[}deg{]} one (Fig. 15 - \emph{second
row}) the \emph{IA-LBE} bounds and the band of the $M=10^{4}$ \emph{RG}
patterns from the \emph{MC} simulations, \{$\Phi\left(\theta,\phi;\underline{p}_{MC}^{\left(m\right)}\right)$;
$m=1,...,M$\}, predicted with the same \emph{EM} full-wave solver,
but considering as input parameters, \{$\underline{p}_{MC}^{\left(m\right)}$;
$m=1,...,M$\}, a set of random samples different from the training
ones {[}i.e., $\underline{p}_{MC}^{\left(m\right)}\in\left[\underline{p}\right]$
and $\underline{p}_{MC}^{\left(m\right)}\neq\underline{p}^{\left(s\right)}$
($m=1,...,M$; $s=1,...,S$){]}. As it can be observed {[}Figs. 15(\emph{a})-15(\emph{e}){]},
the \emph{IA-LBE} bounds are tight and inclusive ($\left.\Psi\right\rfloor _{\phi=0\,[\textnormal{deg}]}=1.55$
and $\left.\Psi\right\rfloor _{\phi=90\,[\textnormal{deg}]}=2.66$
- Tab. VI). They have been derived in $\tau_{IA-LBE}=24$ {[}min{]},
the main amount of time being spent for the full-wave \emph{EM} simulation
of the $S=6$ training samples since the remaining \emph{CPU}-time
for the computation of the \emph{IA} bounds took $1.6\times10^{-2}$
{[}sec{]}. In comparison, the \emph{MC} simulations required $\tau_{MC}=43200$
{[}min{]} (i.e., $\approx30$ days), thus the \emph{IA-LBE} enabled
an overall time saving of $\Delta\tau_{MC}^{IA-LBE}\approx99.94$
\%.

\noindent The same experiment has been repeated by still considering
one (i.e., $N=1$) uncertain parameter, but different, that is either
the patch width ($PW$), or the offset between the slot and the feeding
line stub ($O_{S}$), or the substrates permittivity ($\epsilon_{r}$).
In all cases, the tolerance value has been set to $\delta=3$ \% (Tab.
V). The plots of the \emph{IA-LBE} bounds and the corresponding \emph{MC}
simulations along the principal planes of the pattern (i.e., $\phi=0$
{[}deg{]} {[}Figs. 15(\emph{b})-15(\emph{d}){]} and $\phi=90$ {[}deg{]}
{[}Figs. 15(\emph{f})-15(\emph{h}){]}) are given in Fig. 15(\emph{a})
and Fig. 15(\emph{e}) when $\left.p_{n}\right\rfloor _{n=1}=h_{0}$,
Fig. 15(\emph{b}) and Fig. 15(\emph{f}) when $\left.p_{n}\right\rfloor _{n=1}=WP$,
Fig. 15(\emph{c}) and Fig. 15(\emph{g}) when $\left.p_{n}\right\rfloor _{n=1}=O_{S}$,
and Fig. 15(\emph{d}) and Fig. 15(\emph{h}) when $\left.p_{n}\right\rfloor _{n=1}=\epsilon_{r}$.
For completeness, Table VI gives the corresponding values of $\Psi_{IA-LBE}$,
which are always positive.

\noindent Finally, the same antenna structure has been analyzed when
all previous $N=4$ parameters (i.e., $\underline{p}=\left\{ h_{0},\, WP,\, O_{s},\,\epsilon_{r}\right\} $)
are affected by tolerances. As expected and in line with the results
of the previous examples, even though concerned with different \emph{TF}s,
the \emph{IA} bounds are looser than those for the single ($N=1$)
parameter tests (e.g., $\left.\Psi_{IA-LBE}\right\rfloor _{\phi=0\,[\textnormal{deg}]}=1.51$
and $\left.\Psi_{IA-LBE}\right\rfloor _{\phi=90\,[\textnormal{deg}]}=2.45$
- Tab. VI). More important, once again the \emph{IA-LBE} approach
confirms to be reliable, effective, and computationally efficient
($\tau_{IA-LBE}=96$ {[}min{]} vs. $\tau_{MC}\approx28$ {[}d{]} $\to$
$\Delta\tau_{MC}^{IA-LBE}=99.76$ \%) in providing non-trivial (i.e.,
finite) and inclusive bounds ($\Psi>0$) despite the \emph{TF} is
not available in explicit/closed form, but its {}``knowledge'' is
limited to the availability of some ($S=S_{th}$, $S_{th}=6\times N$)
I/O samples.

\section{\noindent Conclusions\label{sec:Conclusions}}

An innovative analytic method for the sensitivity analysis of \emph{EM}
systems, whose \emph{TF} between the control parameters - affected
by unknown, but bounded, tolerances - and the system performance response
is not available in explicit/closed-form, has been presented. Starting
from a set of available I/O examples, a \emph{SM} of the \emph{TF}
has been defined with a \emph{LBE} approach and it has been then extended
to intervals through \emph{IA} for yielding analytic and inclusive,
yet finite, performance bounds.

\noindent From the numerical assessment, the following main outcomes
can be drawn:

\begin{itemize}
\item the size of training set, $S$, to yield non-trivial, but inclusive
and reliable, bounds should be at least greater than the threshold
value $S_{th}=6\times N$ (i.e., $S\ge S_{th}$);
\item the \emph{IA-LBE} performance bounds prove to be reliable (i.e., inclusive
of all \emph{MC} simulations) and meaningful (i.e., tight to the \emph{MC}
confidence band and, when the \emph{TF} is available as for the validation
checks in Sect. \ref{sec:Numerical-Results}, close to the asymptotic
\emph{IA-E} ones);
\item analogously to the \emph{IA-E} bounds, the \emph{IA-LBE} ones widen
as the number of descriptors of the \emph{EM} device (i.e., here the
antenna device), which deviate from the nominal values, increases
due to the well-known \emph{IA} {}``\emph{wrapping}'' effect.
\end{itemize}
Future research activities, beyond the scope of this work, will be
aimed at extending the proposed \emph{IA-LBE} method to the sensitivity
analysis of other \emph{EM} devices and systems as well as considering
probabilistic interval distributions.

\newpage
\section*{Appendix I\label{sec:Appendix-I}}

The sum of two real-valued intervals $\left[\chi\right]=\left[\inf\left(\chi\right);\sup\left(\chi\right)\right]$
and $\left[\psi\right]=\left[\inf\left(\psi\right);\sup\left(\psi\right)\right]$
is defined as \cite{Moore 1966}\begin{equation}
\left[\chi\right]+\left[\psi\right]=\left[\inf\left(\chi\right)+\inf\left(\psi\right);\sup\left(\chi\right)+\sup\left(\psi\right)\right].\label{eq:_App-I.interval-sum}\end{equation}

\section*{Appendix II\label{sec:Appendix-II}}

Let $\left[\chi\right]=\left[\inf\left(\chi\right);\sup\left(\chi\right)\right]$
be a real-valued interval. The power of $\left[\chi\right]$, $\left[\chi\right]^{\alpha}$,
$\alpha$ being a positive integer, is a real-valued interval defined
as \cite{Moore 1966}\begin{equation}
\left[\chi\right]^{\alpha}=\left\{ \begin{array}{lll}
\left[\left\{ \inf\left(\chi\right)\right\} ^{\alpha};\left\{ \sup\left(\chi\right)\right\} ^{\alpha}\right] & if & \left\{ \inf\left(\chi\right)>0\right\} \,\,|\,\,\left(\alpha\mathrm{mod}2\neq0\right)\\
\left[\left\{ \sup\left(\chi\right)\right\} ^{\alpha};\left\{ \inf\left(\chi\right)\right\} ^{\alpha}\right] & if & \left\{ \sup\left(\chi\right)<0\right\} \,\,\&\,\,\left(\alpha\mathrm{mod}2=0\right)\\
\left[0;\max\left\{ \left\{ \inf\left(\chi\right)\right\} ^{\alpha},\left\{ \sup\left(\chi\right)\right\} ^{\alpha}\right\} \right] & if & 0\in\left[\chi\right]\end{array}\right.\label{eq:_App-I.interval-exp-1}\end{equation}
where $\mathrm{mod}$ is the modulo operation. As the interval of
the absolute value of $\left[\chi\right]$, $\left|\left[\chi\right]\right|$,
is equal to \cite{Moore 1966}\begin{equation}
\left|\left[\chi\right]\right|=\left\{ \begin{array}{lll}
\left[\inf\left(\chi\right);\sup\left(\chi\right)\right] & if & \left\{ \inf\left(\chi\right)>0\right\} >0\\
\left[\sup\left(\chi\right);\inf\left(\chi\right)\right] & if & \left\{ \sup\left(\chi\right)<0\right\} \\
\left[0;\,\max\left\{ \inf\left(\chi\right),\sup\left(\chi\right)\right\} \right] & if & 0\in\left[\chi\right]\end{array}\right.,\label{eq:_App-I.interval-exp-2}\end{equation}
it follows that the power of the absolute value of $\left[\chi\right]$,
$\left|\left[\chi\right]\right|^{\alpha}$, turns out being a real-valued
interval defined as \cite{Moore 1966}\begin{equation}
\left|\left[\chi\right]\right|^{\alpha}=\left\{ \begin{array}{lll}
\left[\left\{ \inf\left(\chi\right)\right\} ^{\alpha};\left\{ \sup\left(\chi\right)\right\} ^{\alpha}\right] & if & \left\{ \inf\left(\chi\right)>0\right\} \\
\left[\left\{ \sup\left(\chi\right)\right\} ^{\alpha};\left\{ \inf\left(\chi\right)\right\} ^{\alpha}\right] & if & \left\{ \sup\left(\chi\right)<0\right\} \\
\left[0;\max\left\{ \left\{ \inf\left(\chi\right)\right\} ^{\alpha},\left\{ \sup\left(\chi\right)\right\} ^{\alpha}\right\} \right] & if & 0\in\left[\chi\right]\end{array}\right..\label{eq:_App-I.interval-exp-3}\end{equation}

\section*{\noindent Acknowledgements\label{sec:Acknowledgements}}

\noindent This work benefited from the networking activities carried
out within the Project {}``AURORA - Smart Materials for Ubiquitous
Energy Harvesting, Storage, and Delivery in Next Generation Sustainable
Environments'' funded by the Italian Ministry for Universities and
Research within the PRIN-PNRR 2022 Program. Moreover, it benefited
from the networking activities carried out within the Project {}``SEME@TN
- Smart ElectroMagnetic Environment in TrentiNo'' funded by the Autonomous
Province of Trento (CUP: C63C22000720003), the Project DICAM-EXC (Grant
L232/2016) funded by the Italian Ministry of Education, Universities
and Research (MUR) within the 'Departments of Excellence 2023-2027'
Program (CUP: E63C22003880001), the Project {}``Telecommunications
of the Future'' (PE00000001 - program {}``RESTART''), funded by
the European Union - Next Generation EU under the Italian National
Recovery and Resilience Plan (NRRP), Mission 4, Component 2, Investment
1.3 (CUP: D43C22003080001; J33C22002880001; B53C22003970001), the
Project {}``National Centre for HPC, Big Data and Quantum Computing
(CN HPC)'' funded by the European Union - NextGenerationEU within
the PNRR Program (CUP: E63C22000970007), and the support of the Natural
Science Basic Research Program of Shaanxi Province (Grants No. 2022-JC-33,
No. 2023-GHZD-35, and No. 2024JC-ZDXM-25). Views and opinions expressed
are however those of the author(s) only and do not necessarily reflect
those of the European Union or the European Research Council. Neither
the European Union nor the granting authority can be held responsible
for them. A. Massa wishes to thank E. Vico and L. Massa for the never-ending
inspiration, support, guidance, and help.

\newpage
\section*{FIGURE CAPTIONS}

\begin{itemize}
\item \textbf{Figure 1.} Sketch of an ideal (i.e., w/o tolerance), $\mathbf{J}\left(\mathbf{r}\,;\underline{p}\right)$,
and non-ideal (i.e., w/ tolerance), $\mathbf{J}\left(\mathbf{r}\,;\underline{p}'\right)$,
input function along with the corresponding system responses, $\Phi\left(\theta;\underline{p}\right)$
and $\Phi\left(\theta;\underline{p}'\right)$, obtained throughout
the \emph{TF}, $\mathcal{F}\left(\underline{p}\right)=\Phi\left(\theta;\underline{p}\right)$.
\item \textbf{Figure 2.} Sketch of a system response (or output) function
in its explicit form, $\Phi\left(\theta;\underline{p}\right)$, the
implicit version {[}i.e., the set of $S$ samples, \{$\Phi^{\left(s\right)}$;
$s=1,...,S$\}, $\Phi^{\left(s\right)}\triangleq\Phi\left(\theta;\underline{p}^{\left(s\right)}\right)${]}
and the corresponding surrogate form, $\widetilde{\Phi}\left(\theta;\underline{p}\right)$.
\item \textbf{Figure 3.} \emph{Benchmark Function} ($N=1$, $\delta=20$
\%) - Plot of the \emph{IA-LBE} bounds from a training set with (\emph{a})
$S=N$, (\emph{b}) $S=2\times N$, (\emph{c}) $S=3\times N$, (\emph{d})
$S=4\times N$, (\emph{e}) $S=5\times N$, (\emph{f}) $S=6\times N$,
(\emph{g}) $S=7\times N$, and (\emph{h}) $S=8\times N$ training
samples together with the nominal explicit function, the $M=10^{6}$
Monte Carlo realizations, the \emph{IA-E} bounds.
\item \textbf{Figure 4.} \emph{Benchmark Function} ($N=1$, $\delta=20$
\%) - Behavior of the \emph{interval inclusion metric}, $\Psi$, versus
the (normalized) number of training samples, $\frac{S}{N}$.
\item \textbf{Figure 5.} \emph{Benchmark Function} ($N=1$, $S=6\times N$)
- Behavior of the \emph{interval inclusion metric}, $\Psi$, versus
the tolerance value, $\delta$.
\item \textbf{Figure 6.} \emph{Benchmark Function} ($N=1$, $S=6\times N$)
- \textbf{}Plot of the $M=10^{6}$ Monte Carlo realizations, the \emph{IA-E}
bounds, and the \emph{IA-LBE} bounds derived when considering (\emph{a})
$\delta=5\%$, (\emph{b}) $\delta=10\%$, (\emph{c}) $\delta=30\%$,
(\emph{d}) $\delta=40\%$, and (\emph{e}) $\delta=50\%$ together
with the nominal explicit function.
\item \textbf{Figure 7.} \emph{Benchmark Function} ($S=6\times N$, $\delta=20$
\%) - Behavior of the \emph{interval inclusion metric}, $\Psi$, versus
the number of descriptive parameters affected by tolerances, $N$.
\item \textbf{Figure 8.} \emph{Benchmark Function} ($S=6\times N$, $\delta=20$
\%) - Plot of the nominal explicit function, the $M=10^{6}$ Monte
Carlo realizations, the \emph{IA-E} bounds, and the \emph{IA-LBE}
bounds when (\emph{a}) $N=2$, (\emph{b}) $N=3$, (\emph{c}) $N=4$,
and (\emph{d}) $N=10$.
\item \textbf{Figure 9.} \emph{Antenna Array} ($N=10$, $d=\frac{\lambda}{2}$,
$SLL=-20$ {[}dB{]}, $S=6\times N$) - Plot of the nominal amplitudes
and intervals of the amplitude of the excitations affected by a tolerance
$\delta$ ($\delta=\{1$ \%, $5$ \%, $10$ \%\}).
\item \textbf{Figure 10.} \emph{Antenna Array} ($N=10$, $d=\frac{\lambda}{2}$,
$SLL=-20$ {[}dB{]}, $S=6\times N$) - Behavior of the \emph{interval
inclusion metric}, $\Psi$, versus the tolerance value, $\delta$.
\item \textbf{Figure 11.} \emph{Antenna Array} ($N=10$, $d=\frac{\lambda}{2}$,
$SLL=-20$ {[}dB{]}, $S=6\times N$) - Plot of the $M=10^{6}$ Monte
Carlo pattern realizations, the \emph{IA-E} pattern bounds, and the
\emph{IA-LBE} pattern bounds when (\emph{a}) $\delta=1\%$, (\emph{b})
$\delta=5\%$, and (\emph{c}) $\delta=10\%$ together with the nominal
explicit pattern function.
\item \textbf{Figure 12.} \emph{Antenna Array} ($d=\frac{\lambda}{2}$,
$SLL=-20$ {[}dB{]}, $S=6\times N$, $\delta=10$ \%) - Plot of (\emph{a})(\emph{b})
the nominal amplitudes and the intervals of the amplitude of the excitations
and (\emph{c})(\emph{d}) the nominal explicit pattern function, the
$M=10^{6}$ Monte Carlo pattern realizations, the \emph{IA-E} pattern
bounds, and the \emph{IA-LBE} pattern bounds when (\emph{a})(\emph{c})
$N=20$ and (\emph{b})(\emph{d}) $N=50$.
\item \textbf{Figure 13.} \emph{Antenna Array} ($d=\frac{\lambda}{2}$,
$SLL=-20$ {[}dB{]}, $S=6\times N$, $\delta=10$ \%) - Behavior of
the \emph{interval inclusion metric}, $\Psi$, versus the number of
array elements affected by a $\delta$ tolerance, $N$.
\item \textbf{Figure 14.} \emph{Patch Antenna} - Sketch of (\emph{a}) the
exploded, (\emph{b}) the top, and (\emph{c}) the side views of the
aperture-coupled stacked square patch antenna \cite{Targonski 1998}.
\item \textbf{Figure 15.} \emph{Patch Antenna} ($N=1$, $S=6\times N$,
$\delta=3$ \%) \emph{-} Plot along (\emph{a})-(\emph{d}) the $\phi=0$
{[}deg{]} and (\emph{e})-(\emph{h}) the $\phi=90$ {[}deg{]} cuts
of the $M=10^{4}$ Monte Carlo patterns and the \emph{IA-LBE} pattern
bounds when (\emph{a})(\emph{e}) $\left.p_{n}\right\rfloor _{n=1}=h_{0}$,
(\emph{b})(\emph{f}) $\left.p_{n}\right\rfloor _{n=1}=WP$, (\emph{c})(\emph{g})
$\left.p_{n}\right\rfloor _{n=1}=O_{S}$, and (\emph{d})(\emph{h})
$\left.p_{n}\right\rfloor _{n=1}=\epsilon_{r}$.
\item \textbf{Figure 16.} \emph{Patch Antenna} ($N=4$, $S=6\times N$,
$\delta=3$ \%) \emph{-} Plot along (\emph{a}) the $\phi=0$ {[}deg{]}
and (\emph{b}) the $\phi=90$ {[}deg{]} cuts of the $M=10^{4}$ Monte
Carlo patterns and the \emph{IA-LBE} pattern bounds when $\underline{p}=\left\{ h_{0},\, WP,\, O_{s},\,\epsilon_{r}\right\} $.
\end{itemize}

\section*{TABLE CAPTIONS}

\begin{itemize}
\item \textbf{Table I.} \emph{Benchmark Function} ($N=1$, $\delta=20$
\%) - Values of the \emph{interval inclusion metric}, $\Psi$, and
the corresponding terms in (\ref{eq:_interval-inclusion-metric}).
\item \textbf{Table II.} \emph{Antenna Array} ($N=10$, $d=\frac{\lambda}{2}$,
$SLL=-20$ {[}dB{]}, $S=6\times N$) - Values of the pattern features,
the \emph{tolerance index}, $\Delta$, and the \emph{interval inclusion
metric}, $\Psi$.
\item \textbf{Table III.} \emph{Antenna Array} ($d=\frac{\lambda}{2}$,
$SLL=-20$ {[}dB{]}, $S=6\times N$, $\delta=10$ \%) - Values of
the pattern features, the \emph{tolerance index}, $\Delta$, and the
\emph{interval inclusion metric}, $\Psi$.
\item \textbf{Table IV.} \emph{Patch Antenna} - Values of the descriptive
parameters.
\item \textbf{Table V.} \emph{Patch Antenna} \emph{-} Tolerance values and
intervals of the descriptive parameters.
\item \textbf{Table VI.} \emph{Patch Antenna} ($S=6\times N$, $\delta=3$
\%) \emph{-} Values of the \emph{interval inclusion metric}, $\Psi$.
\end{itemize}
\newpage
\begin{center}~\vfill\end{center}

\begin{center}\begin{tabular}{c}
\includegraphics[%
  width=0.80\columnwidth]{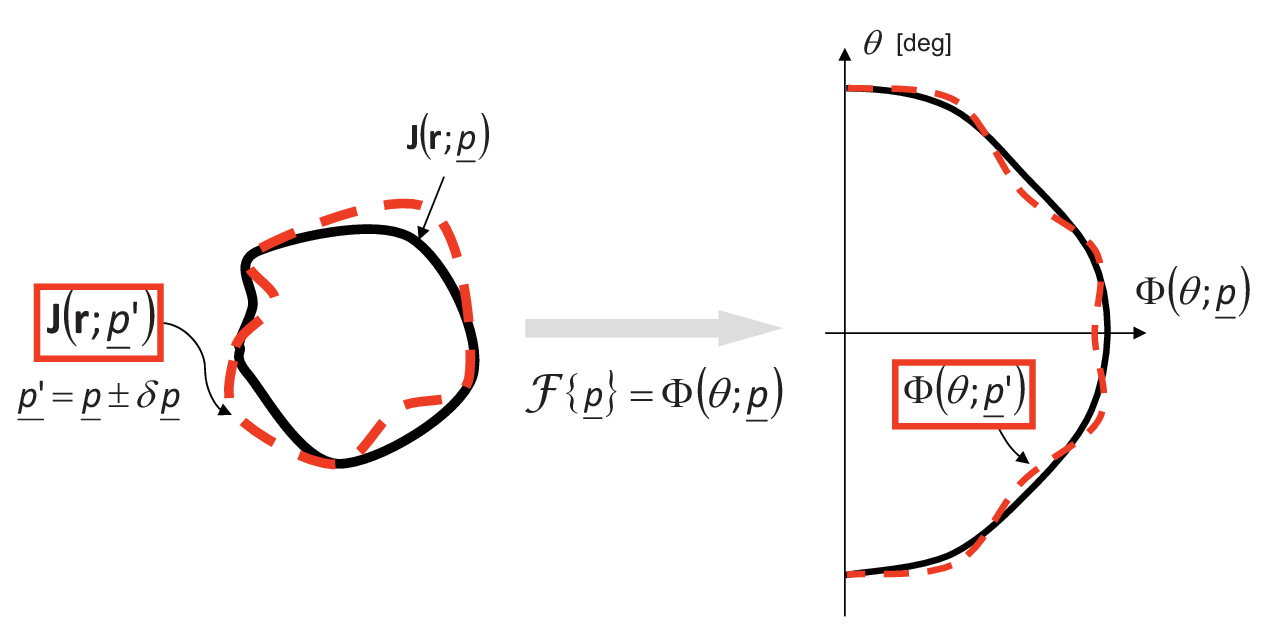}\tabularnewline
\end{tabular}\end{center}

\begin{center}~\vfill\end{center}

\begin{center}\textbf{Fig. 1 - N. Anselmi} \textbf{\emph{et al.}}\textbf{,}
\textbf{\emph{{}``}}Sensitivity Analysis for Antenna Devices ...''\end{center}

\newpage
\begin{center}~\vfill\end{center}

\begin{center}\begin{tabular}{c}
\includegraphics[%
  width=0.70\columnwidth]{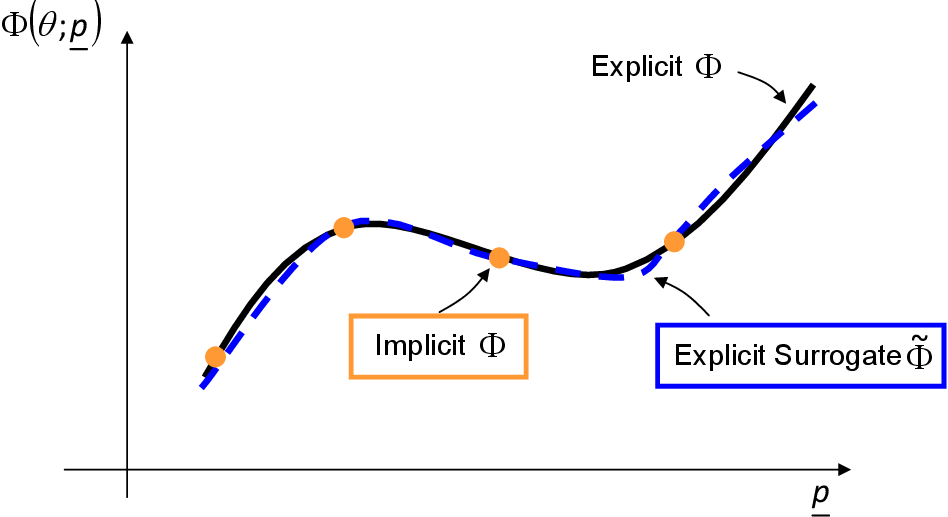}\tabularnewline
\end{tabular}\end{center}

\begin{center}~\vfill\end{center}

\begin{center}\textbf{Fig. 2 - N. Anselmi} \textbf{\emph{et al.}}\textbf{,}
\textbf{\emph{{}``}}Sensitivity Analysis for Antenna Devices ...''\end{center}

\newpage
\begin{center}~\vfill\end{center}

\begin{center}\begin{tabular}{cc}
\includegraphics[%
  width=0.40\columnwidth]{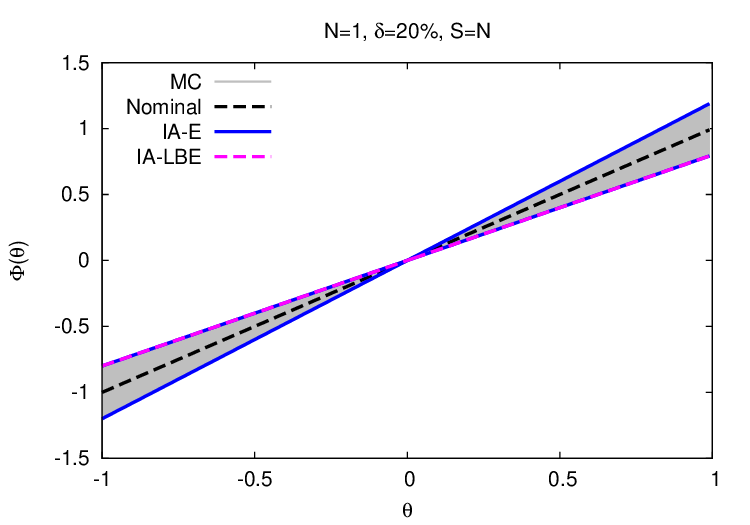}&
\includegraphics[%
  width=0.40\columnwidth]{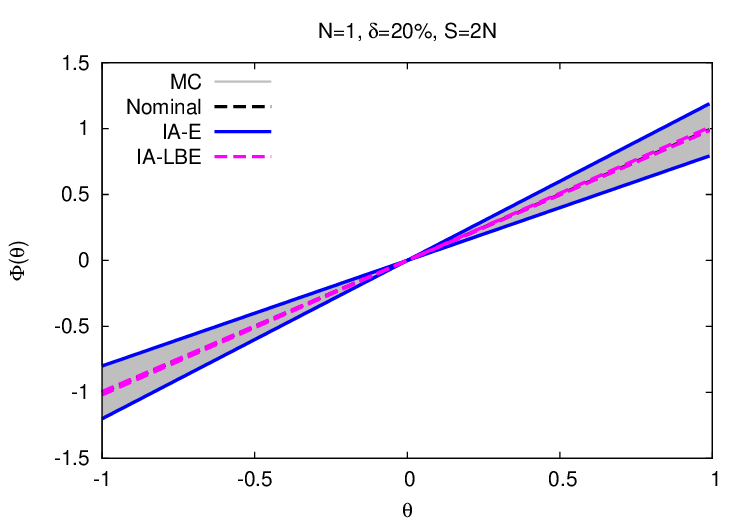}\tabularnewline
(\emph{a})&
(\emph{b})\tabularnewline
\includegraphics[%
  width=0.40\columnwidth]{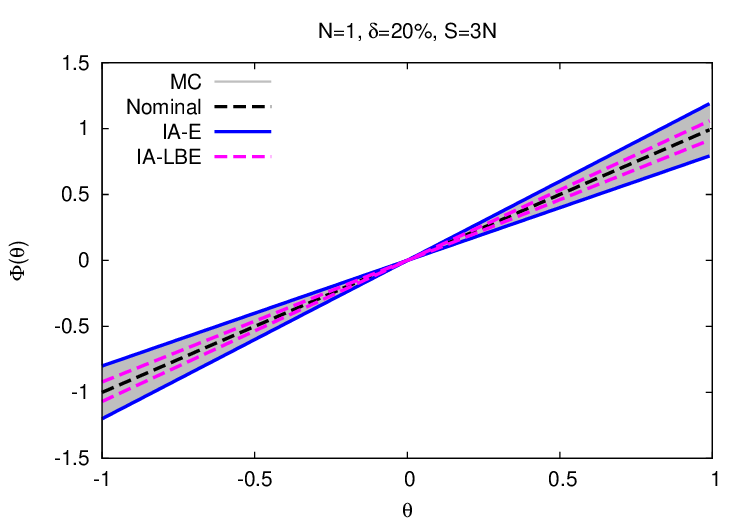}&
\includegraphics[%
  width=0.40\columnwidth]{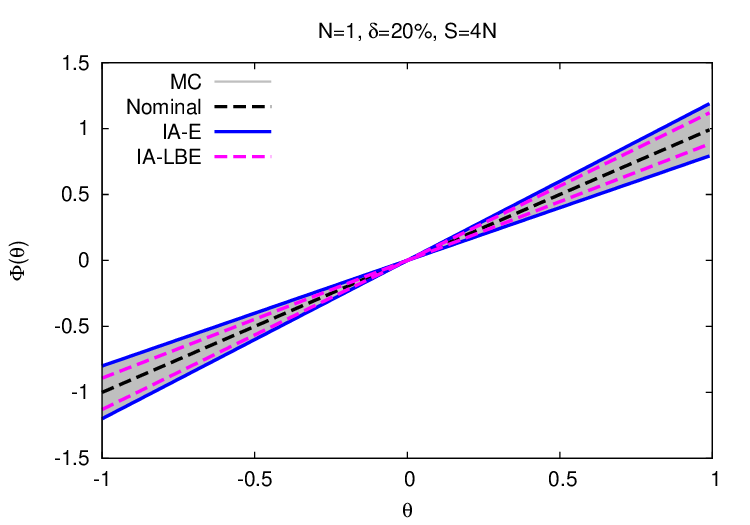}\tabularnewline
(\emph{c})&
(\emph{d})\tabularnewline
\includegraphics[%
  width=0.40\columnwidth]{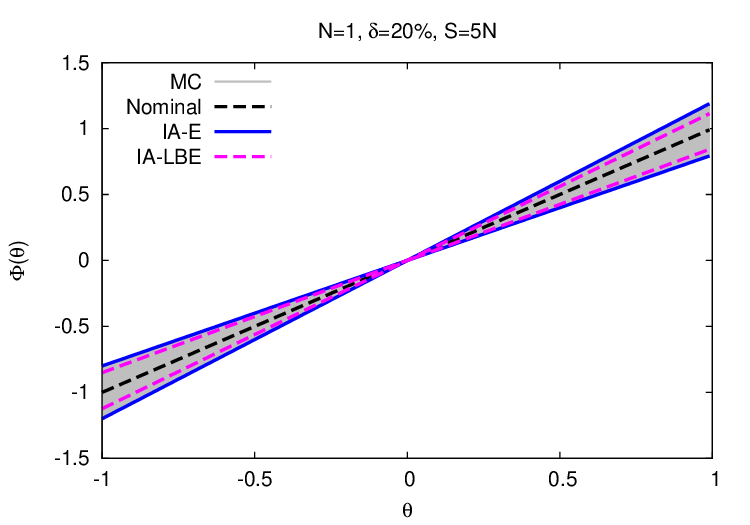}&
\includegraphics[%
  width=0.40\columnwidth]{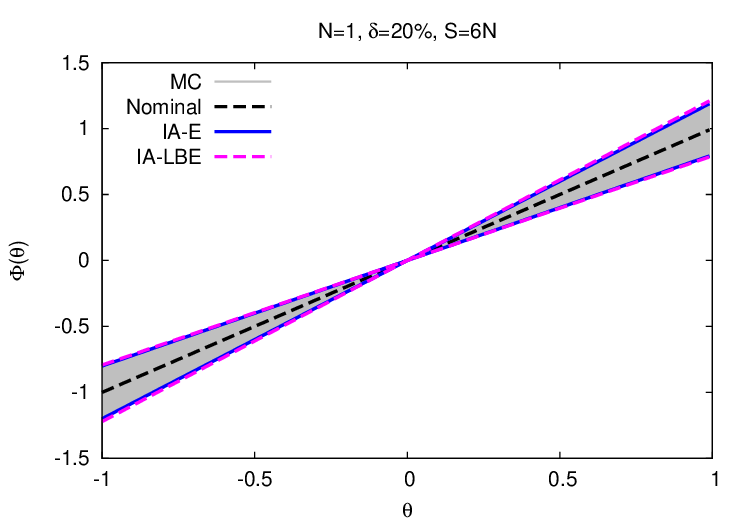}\tabularnewline
(\emph{e})&
(\emph{f})\tabularnewline
\includegraphics[%
  width=0.40\columnwidth]{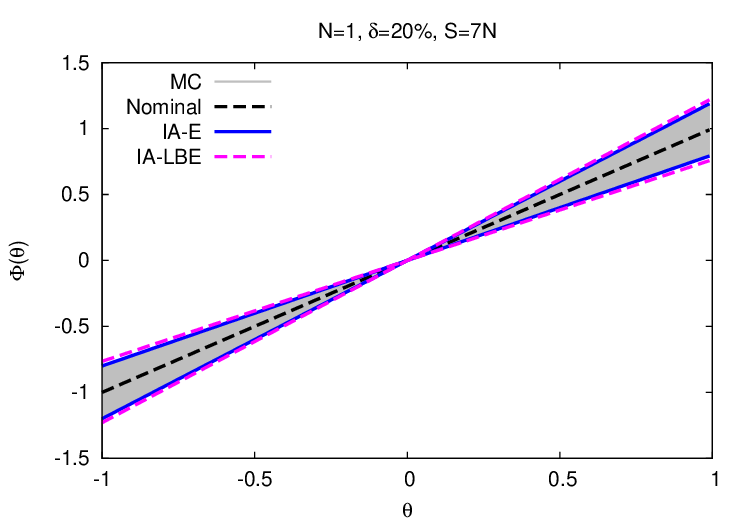}&
\includegraphics[%
  width=0.40\columnwidth]{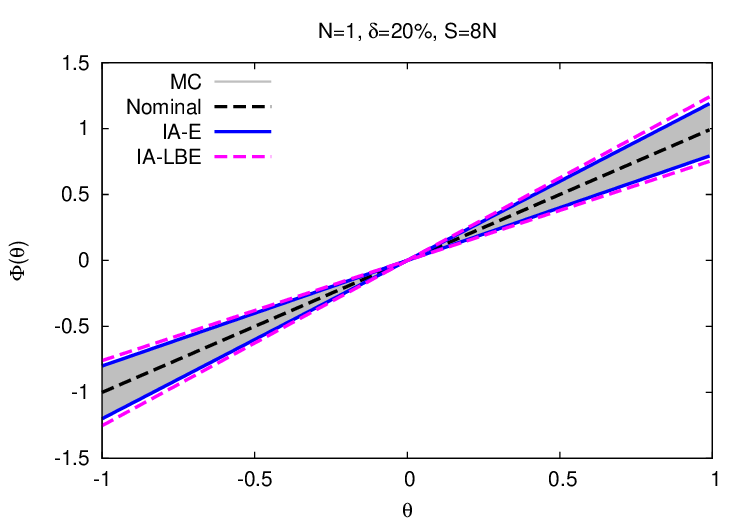}\tabularnewline
(\emph{g})&
(\emph{h})\tabularnewline
\end{tabular}\end{center}

\begin{center}~\vfill\end{center}

\begin{center}\textbf{Fig. 3 - N. Anselmi} \textbf{\emph{et al.}}\textbf{,}
\textbf{\emph{{}``}}Sensitivity Analysis for Antenna Devices ...''\end{center}

\newpage
\begin{center}~\vfill\end{center}

\begin{center}\begin{tabular}{c}
\includegraphics[%
  width=0.70\columnwidth]{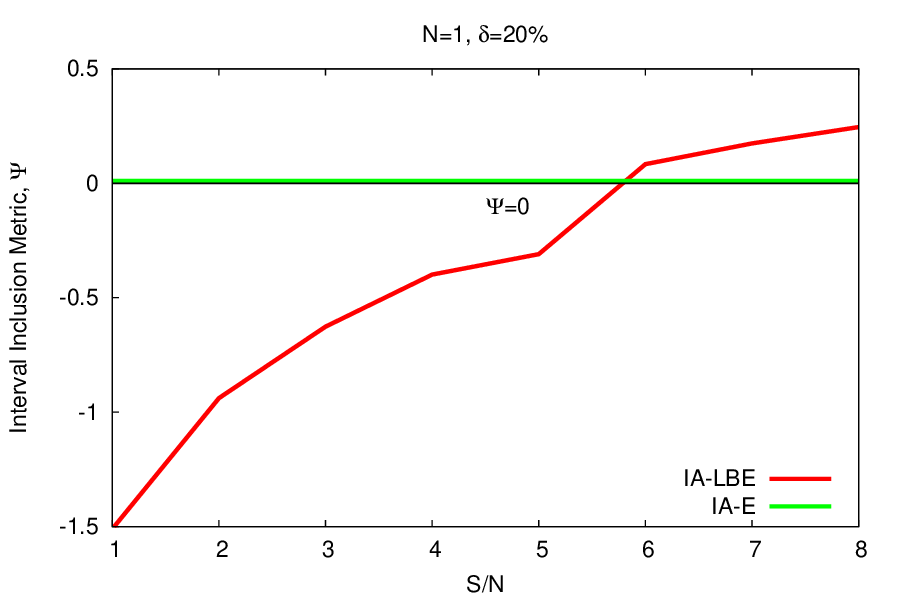}\tabularnewline
\end{tabular}\end{center}

\begin{center}~\vfill\end{center}

\begin{center}\textbf{Fig. 4 - N. Anselmi} \textbf{\emph{et al.}}\textbf{,}
\textbf{\emph{{}``}}Sensitivity Analysis for Antenna Devices ...''\end{center}

\newpage
\begin{center}~\vfill\end{center}

\begin{center}\begin{tabular}{c}
\includegraphics[%
  width=0.70\columnwidth]{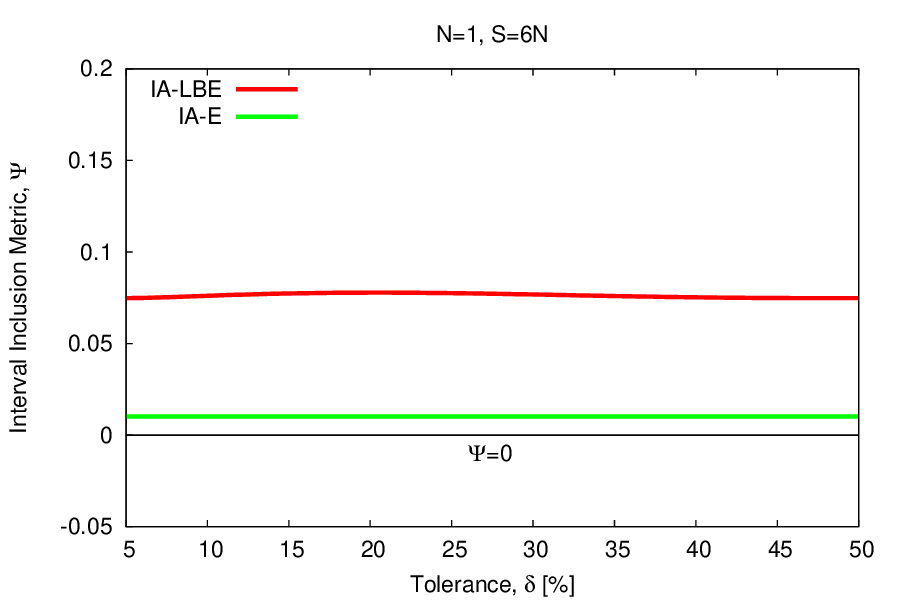}\tabularnewline
\end{tabular}\end{center}

\begin{center}~\vfill\end{center}

\begin{center}\textbf{Fig. 5 - N. Anselmi} \textbf{\emph{et al.}}\textbf{,}
\textbf{\emph{{}``}}Sensitivity Analysis for Antenna Devices ...''\end{center}

\newpage
\begin{center}~\vfill\end{center}

\begin{center}\begin{tabular}{cc}
\includegraphics[%
  width=0.45\columnwidth]{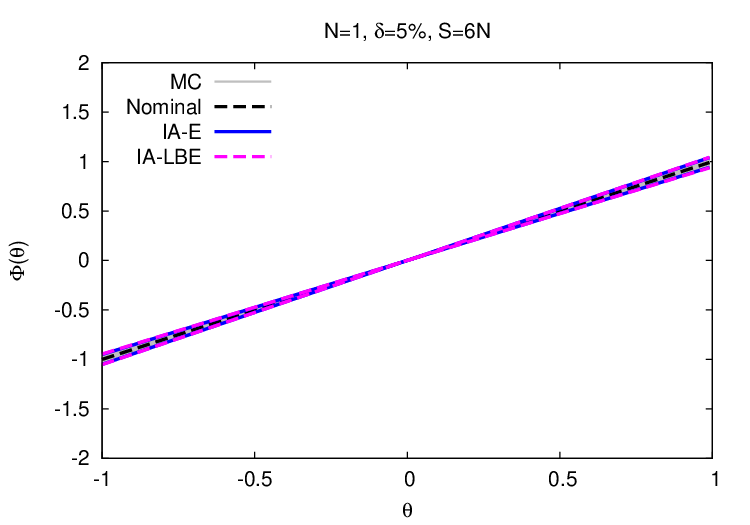}&
\includegraphics[%
  width=0.45\columnwidth]{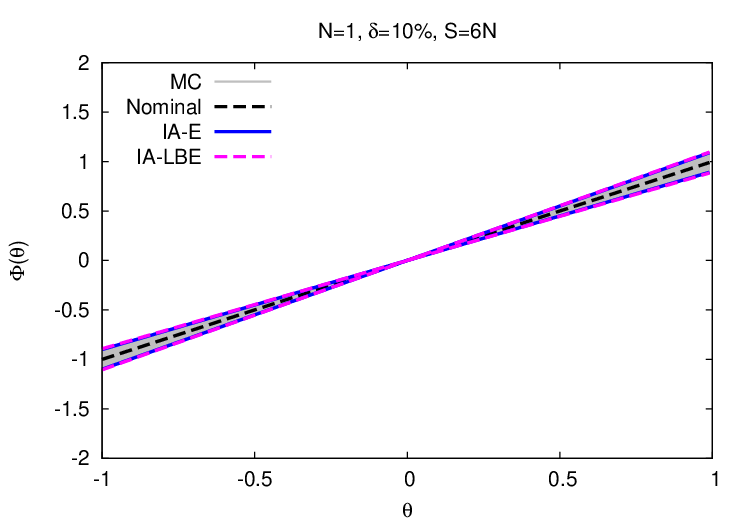}\tabularnewline
(\emph{a})&
(\emph{b})\tabularnewline
\includegraphics[%
  width=0.45\columnwidth]{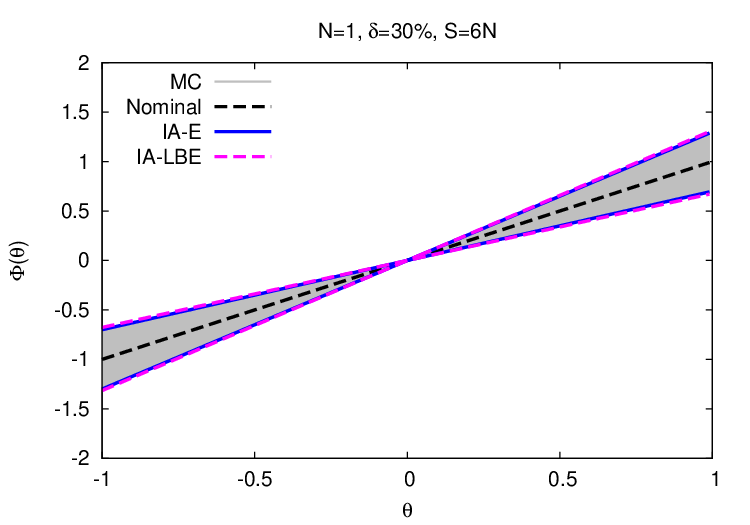}&
\includegraphics[%
  width=0.45\columnwidth]{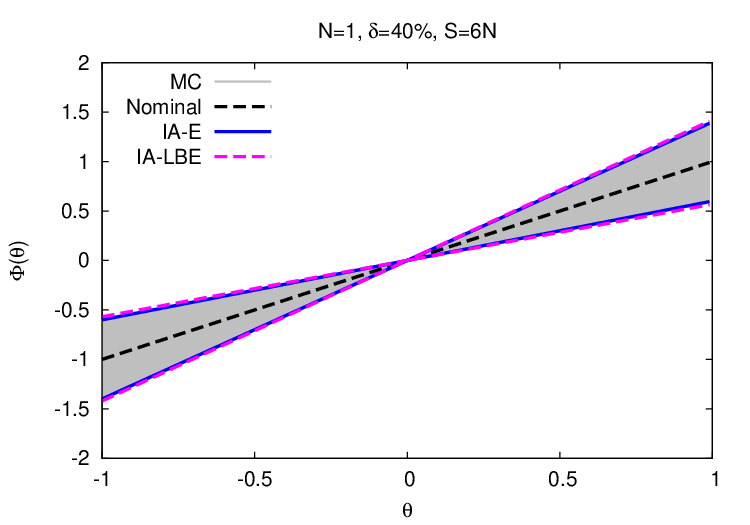}\tabularnewline
(\emph{c})&
(\emph{d})\tabularnewline
\multicolumn{2}{c}{\includegraphics[%
  width=0.45\columnwidth]{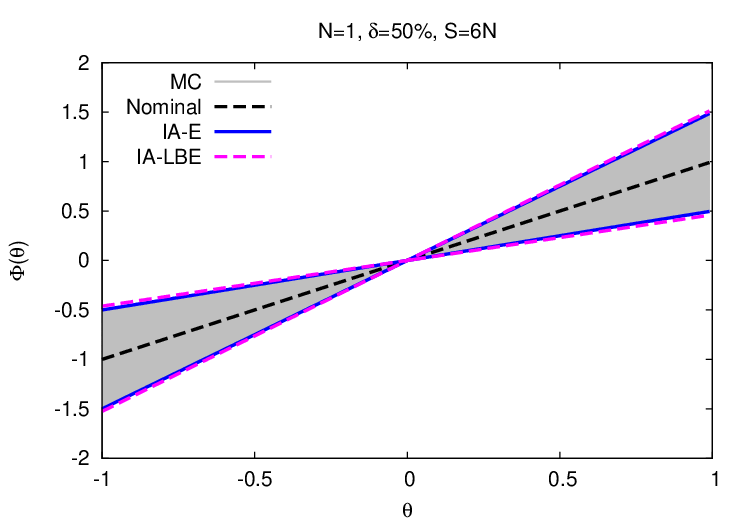}}\tabularnewline
\multicolumn{2}{c}{(\emph{e})}\tabularnewline
\end{tabular}\end{center}

\begin{center}~\vfill\end{center}

\begin{center}\textbf{Fig. 6 - N. Anselmi} \textbf{\emph{et al.}}\textbf{,}
\textbf{\emph{{}``}}Sensitivity Analysis for Antenna Devices ...''\end{center}

\newpage
\begin{center}~\vfill\end{center}

\begin{center}\begin{tabular}{c}
\includegraphics[%
  width=0.70\columnwidth]{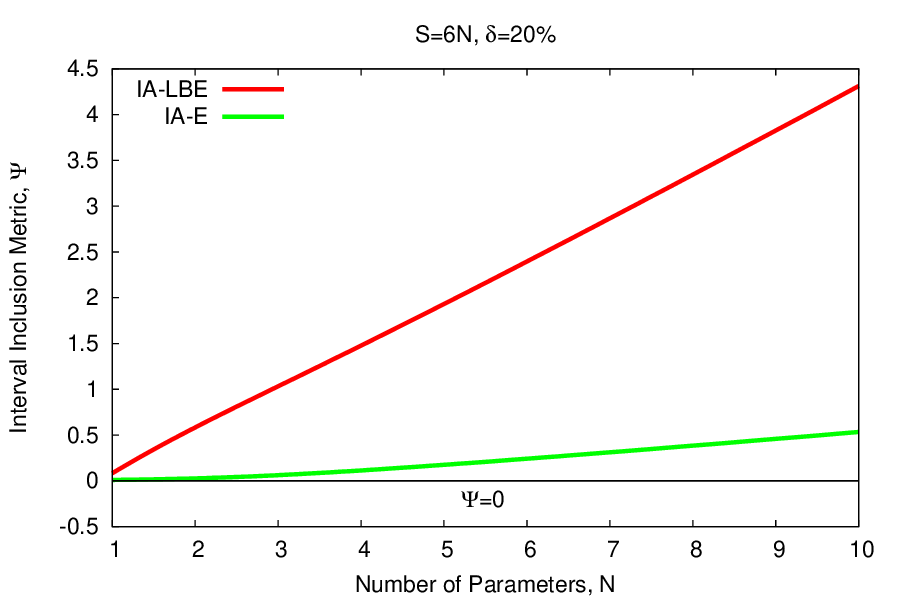}\tabularnewline
\end{tabular}\end{center}

\begin{center}~\vfill\end{center}

\begin{center}\textbf{Fig. 7 - N. Anselmi} \textbf{\emph{et al.}}\textbf{,}
\textbf{\emph{{}``}}Sensitivity Analysis for Antenna Devices ...''\end{center}

\newpage
\begin{center}~\vfill\end{center}

\begin{center}\begin{tabular}{cc}
\includegraphics[%
  width=0.45\columnwidth]{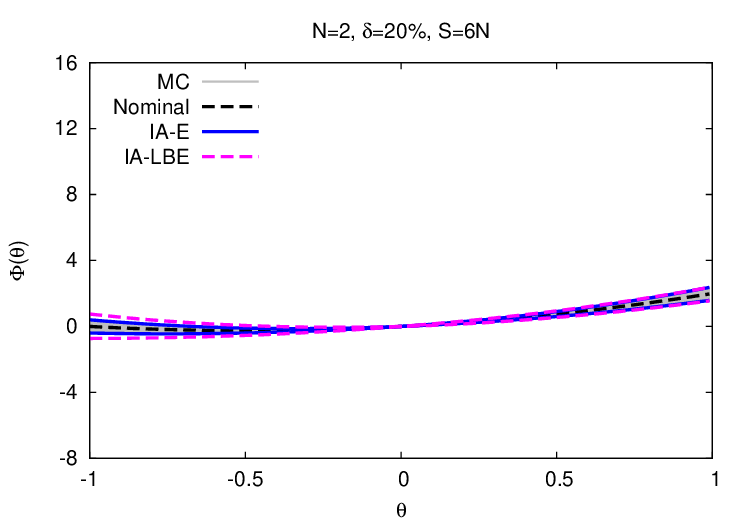}&
\includegraphics[%
  width=0.45\columnwidth]{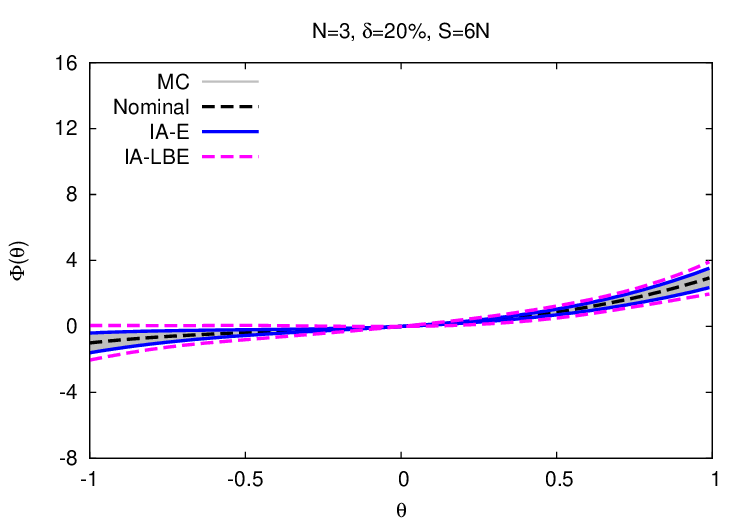}\tabularnewline
(\emph{a})&
(\emph{b})\tabularnewline
\includegraphics[%
  width=0.45\columnwidth]{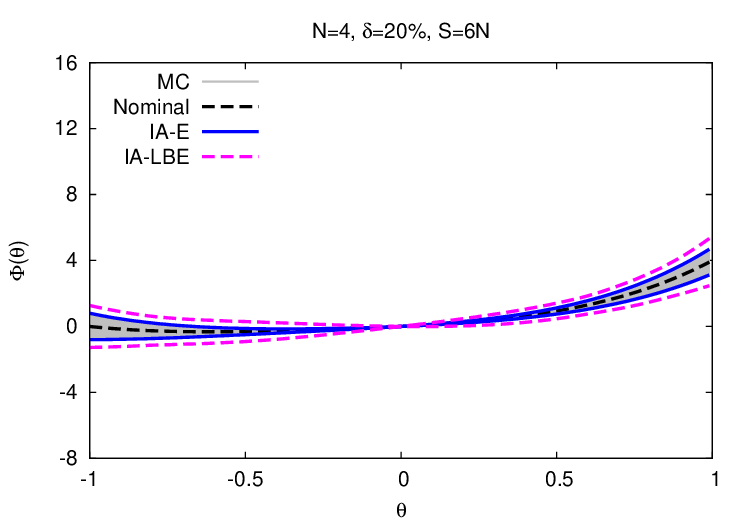}&
\includegraphics[%
  width=0.45\columnwidth]{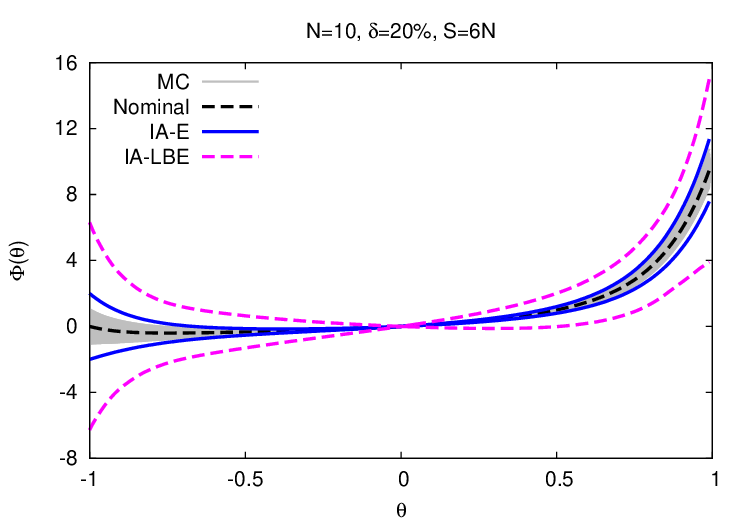}\tabularnewline
(\emph{c})&
(\emph{d})\tabularnewline
\end{tabular}\end{center}

\begin{center}~\vfill\end{center}

\begin{center}\textbf{Fig. 8 - N. Anselmi} \textbf{\emph{et al.}}\textbf{,}
\textbf{\emph{{}``}}Sensitivity Analysis for Antenna Devices ...''\end{center}

\newpage
\begin{center}~\vfill\end{center}

\begin{center}\begin{tabular}{c}
\includegraphics[%
  width=0.70\columnwidth]{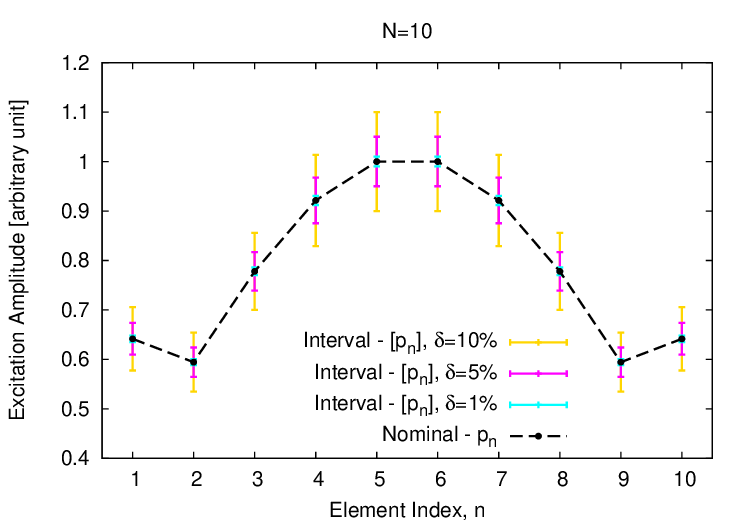}\tabularnewline
\end{tabular}\end{center}

\begin{center}~\vfill\end{center}

\begin{center}\textbf{Fig. 9 - N. Anselmi} \textbf{\emph{et al.}}\textbf{,}
\textbf{\emph{{}``}}Sensitivity Analysis for Antenna Devices ...''\end{center}

\newpage
\begin{center}~\vfill\end{center}

\begin{center}\begin{tabular}{c}
\includegraphics[%
  width=0.70\columnwidth]{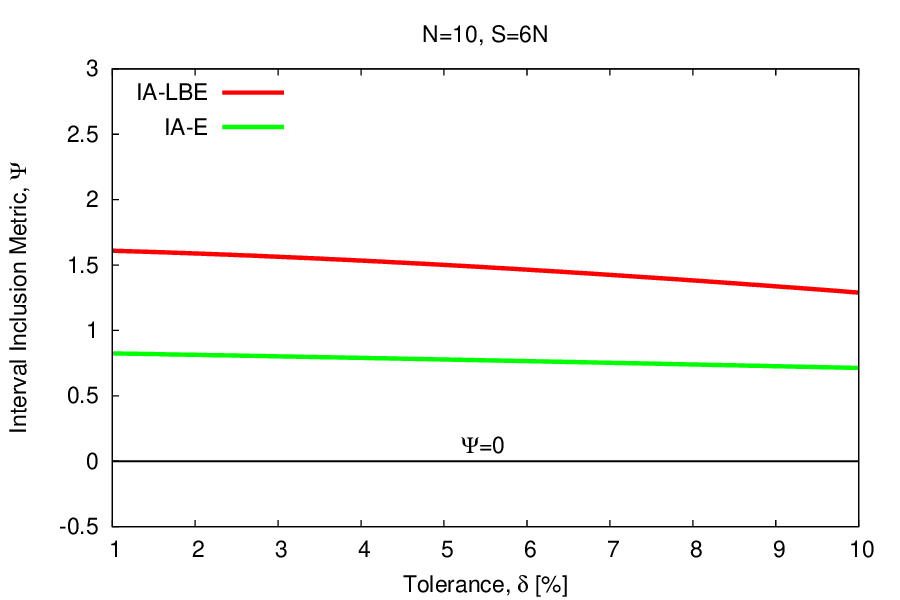}\tabularnewline
\end{tabular}\end{center}

\begin{center}~\vfill\end{center}

\begin{center}\textbf{Fig. 10 - N. Anselmi} \textbf{\emph{et al.}}\textbf{,}
\textbf{\emph{{}``}}Sensitivity Analysis for Antenna Devices ...''\end{center}

\newpage
\begin{center}~\vfill\end{center}

\begin{center}\begin{tabular}{c}
\includegraphics[%
  width=0.50\columnwidth]{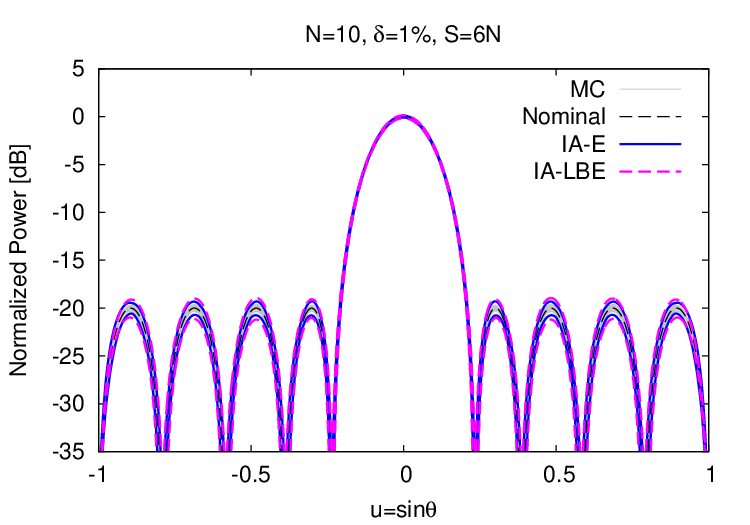}\tabularnewline
(\emph{a})\tabularnewline
\multicolumn{1}{c}{\includegraphics[%
  width=0.50\columnwidth]{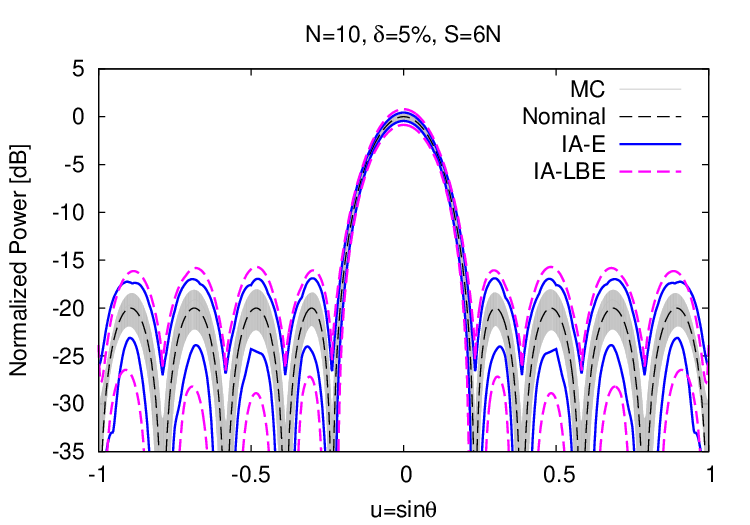}}\tabularnewline
\multicolumn{1}{c}{(\emph{b})}\tabularnewline
\multicolumn{1}{c}{\includegraphics[%
  width=0.50\columnwidth]{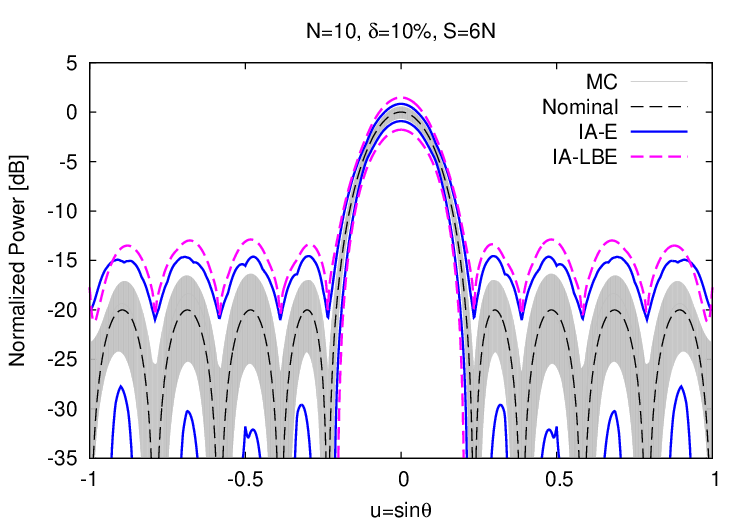}}\tabularnewline
\multicolumn{1}{c}{(\emph{c})}\tabularnewline
\end{tabular}\end{center}

\begin{center}~\vfill\end{center}

\begin{center}\textbf{Fig. 11 - N. Anselmi} \textbf{\emph{et al.}}\textbf{,}
\textbf{\emph{{}``}}Sensitivity Analysis for Antenna Devices ...''\end{center}

\newpage
\begin{center}~\vfill\end{center}

\begin{center}\begin{tabular}{cc}
\includegraphics[%
  width=0.48\columnwidth]{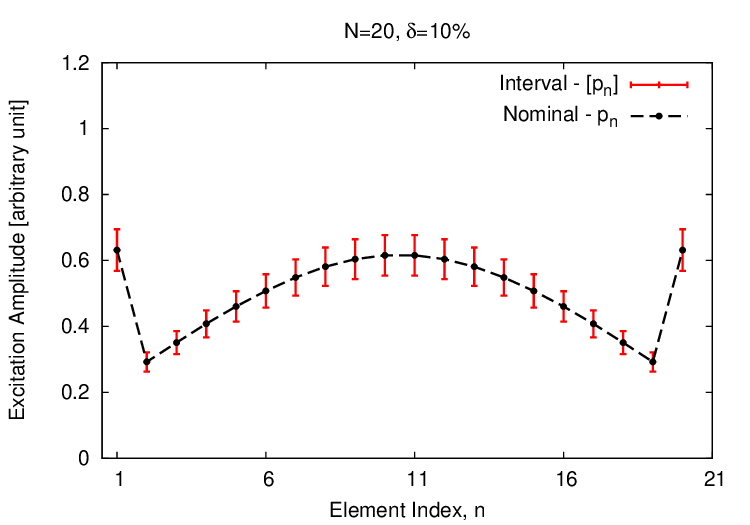}&
\includegraphics[%
  width=0.48\columnwidth]{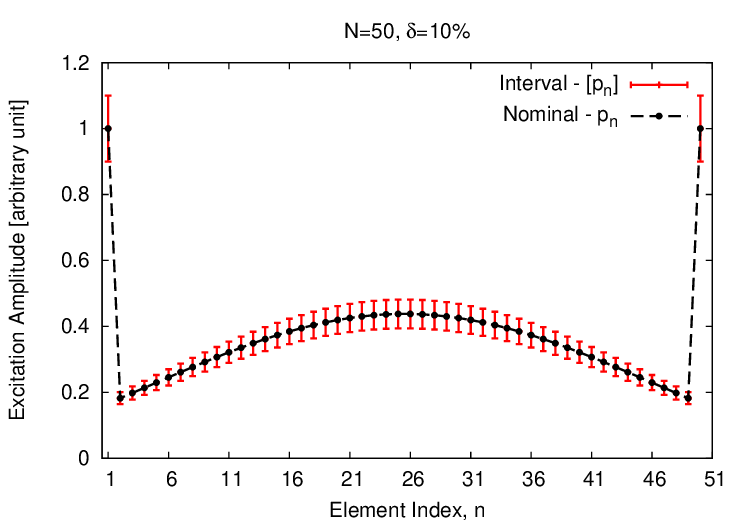}\tabularnewline
(\emph{a})&
(\emph{b})\tabularnewline
\includegraphics[%
  width=0.48\columnwidth]{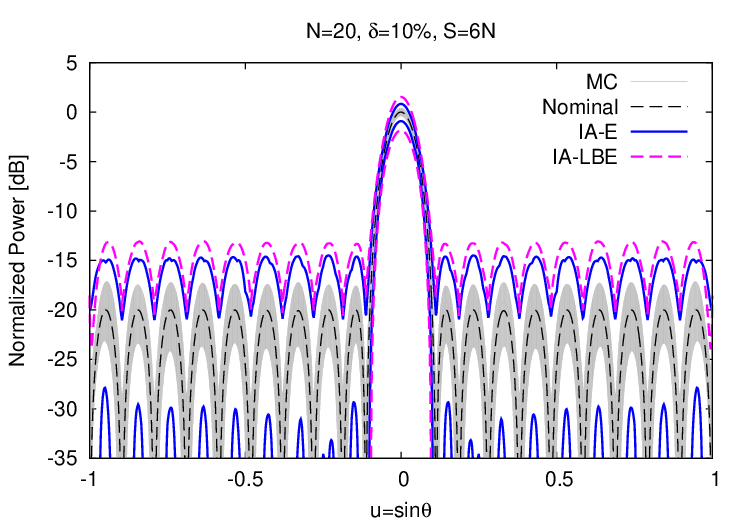}&
\includegraphics[%
  width=0.48\columnwidth]{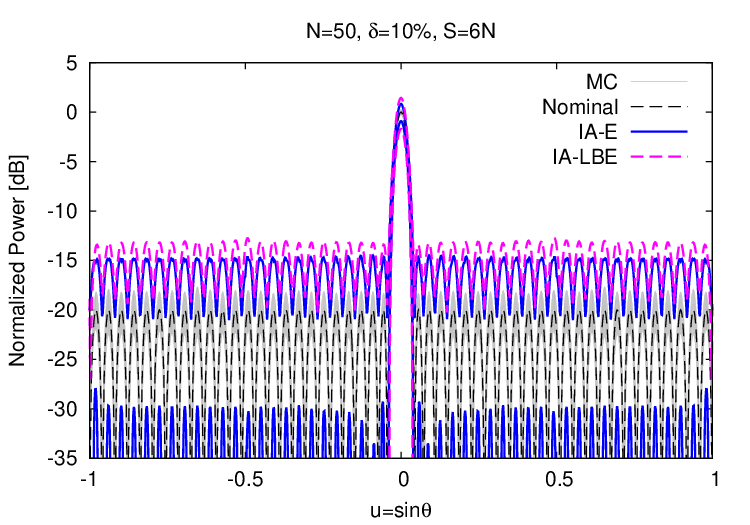}\tabularnewline
(\emph{c})&
(\emph{d})\tabularnewline
\end{tabular}\end{center}

\begin{center}~\vfill\end{center}

\begin{center}\textbf{Fig. 12 - N. Anselmi} \textbf{\emph{et al.}}\textbf{,}
\textbf{\emph{{}``}}Sensitivity Analysis for Antenna Devices ...''\end{center}

\newpage
\begin{center}~\vfill\end{center}

\begin{center}\begin{tabular}{c}
\includegraphics[%
  width=0.70\columnwidth]{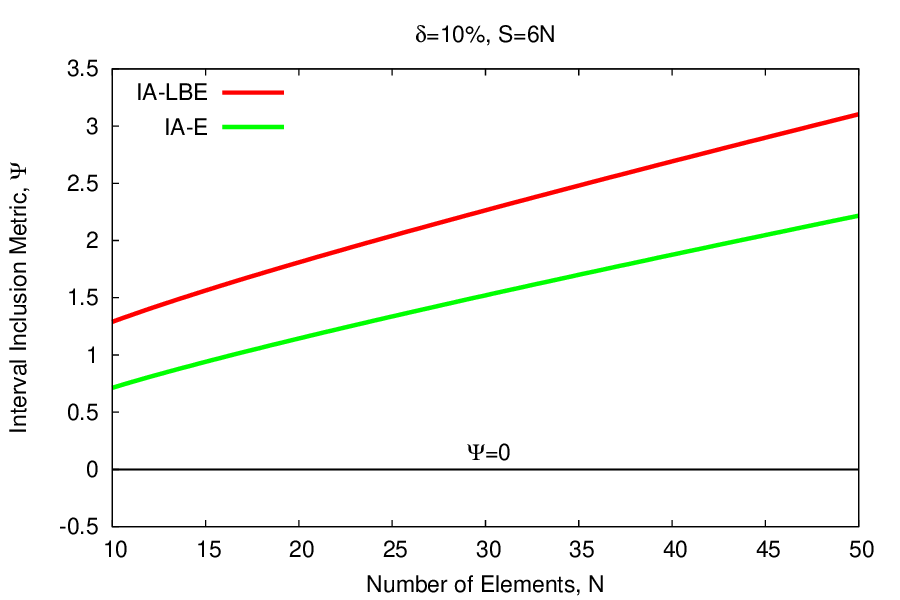}\tabularnewline
\end{tabular}\end{center}

\begin{center}~\vfill\end{center}

\begin{center}\textbf{Fig. 13 - N. Anselmi} \textbf{\emph{et al.}}\textbf{,}
\textbf{\emph{{}``}}Sensitivity Analysis for Antenna Devices ...''\end{center}

\newpage
\begin{center}~\vfill\end{center}

\begin{center}\begin{tabular}{cc}
\multicolumn{2}{c}{\includegraphics[%
  width=0.70\columnwidth]{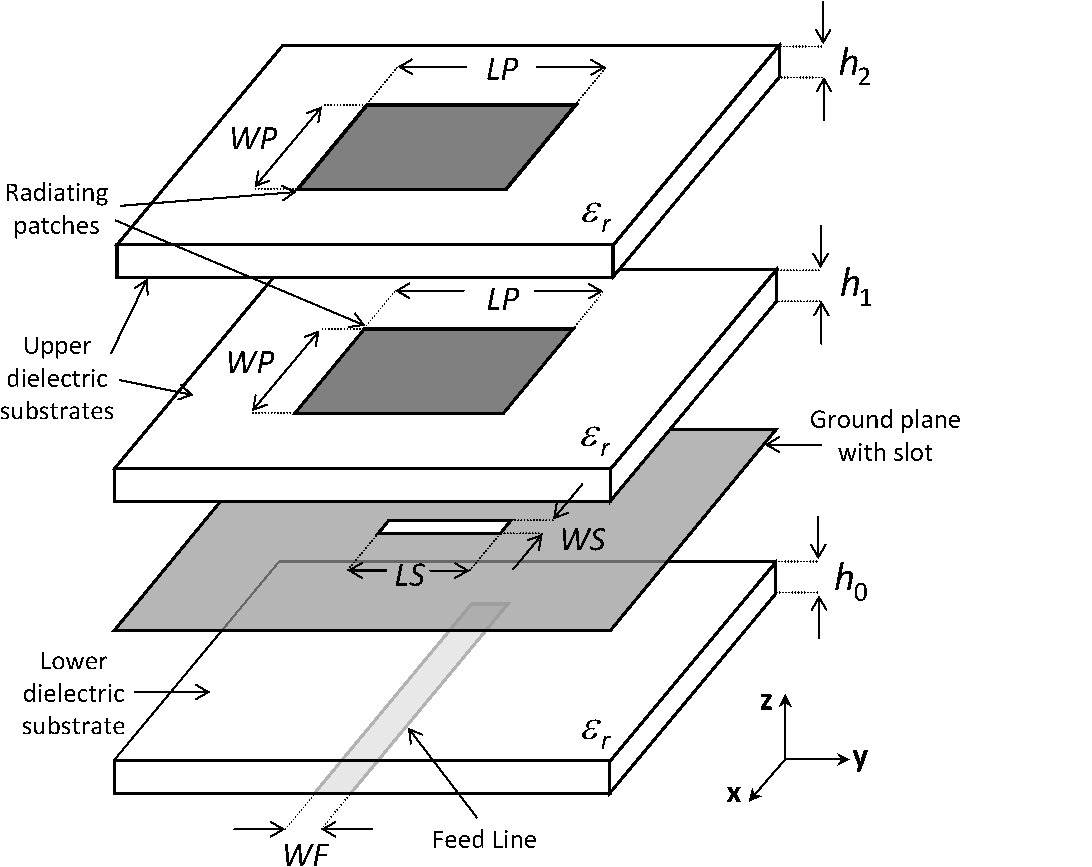}}\tabularnewline
\multicolumn{2}{c}{(\emph{a})}\tabularnewline
\includegraphics[%
  width=0.30\columnwidth]{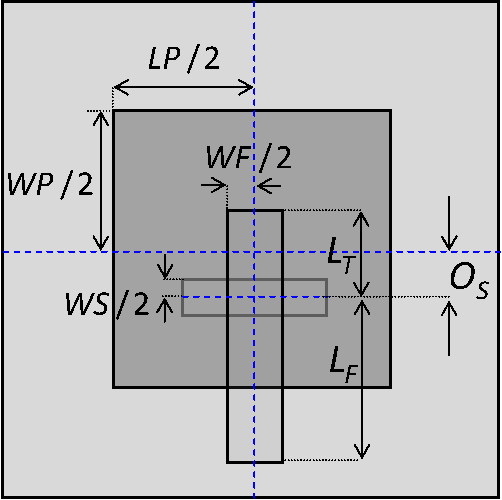}&
\includegraphics[%
  width=0.50\columnwidth]{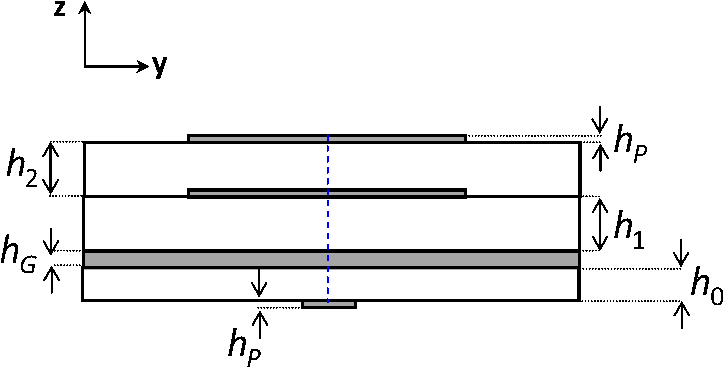}\tabularnewline
(\emph{b})&
(\emph{c})\tabularnewline
\end{tabular}\end{center}

\begin{center}~\vfill\end{center}

\begin{center}\textbf{Fig. 14 - N. Anselmi} \textbf{\emph{et al.}}\textbf{,}
\textbf{\emph{{}``}}Sensitivity Analysis for Antenna Devices ...''\end{center}

\newpage
\begin{center}\begin{sideways}
\begin{tabular}{cccc}
\includegraphics[%
  width=0.33\columnwidth]{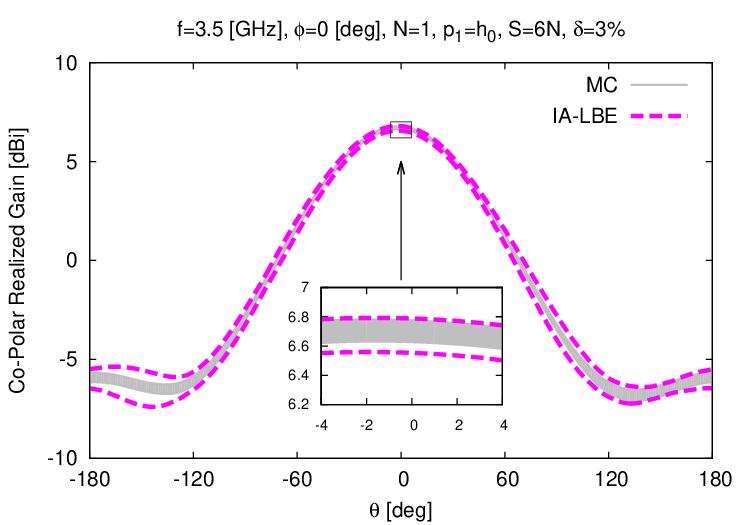}&
\includegraphics[%
  width=0.33\columnwidth]{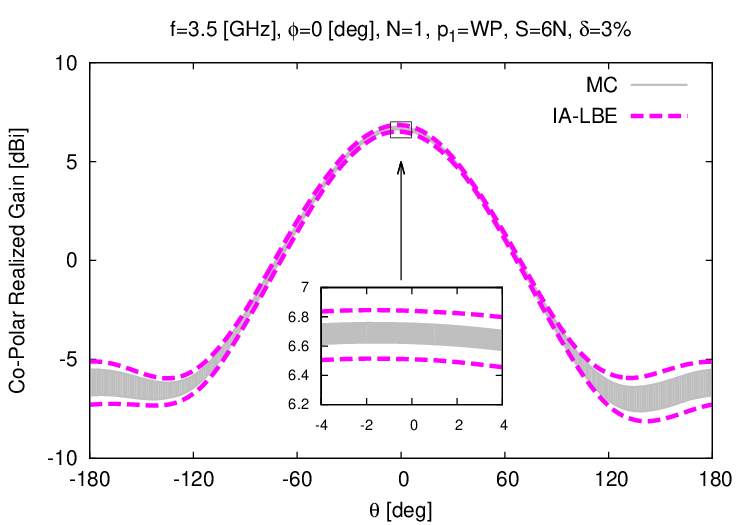}&
\includegraphics[%
  width=0.33\columnwidth]{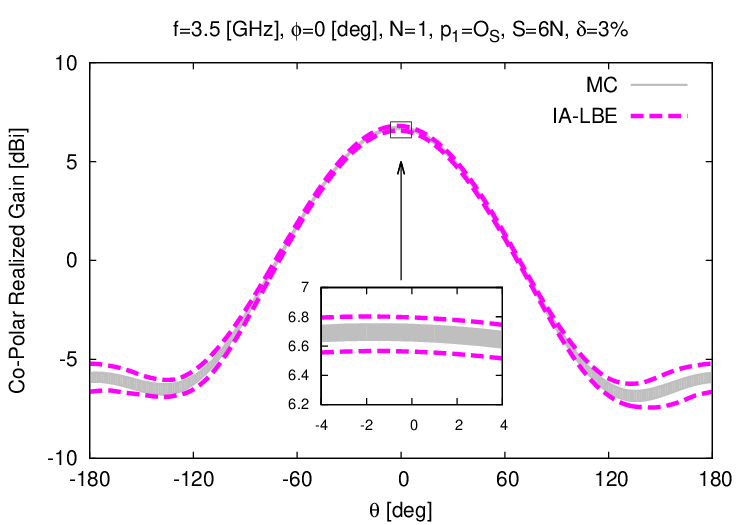}&
\includegraphics[%
  width=0.33\columnwidth]{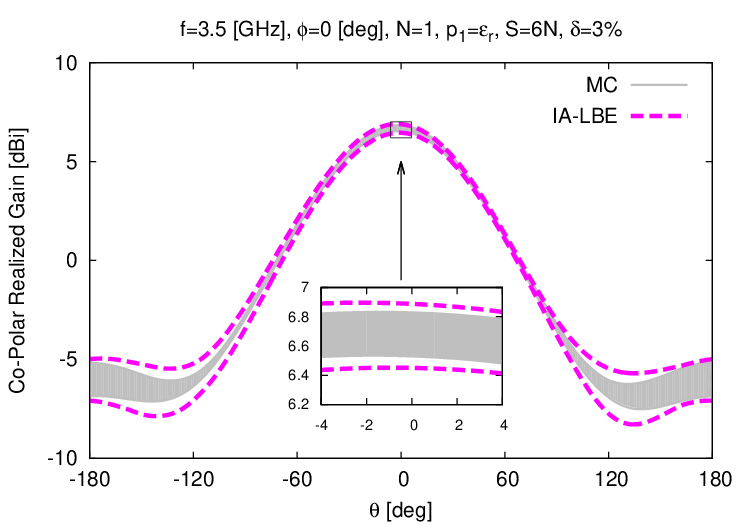}\tabularnewline
(\emph{a})&
(\emph{b})&
(\emph{c})&
(\emph{d})\tabularnewline
\includegraphics[%
  width=0.33\columnwidth]{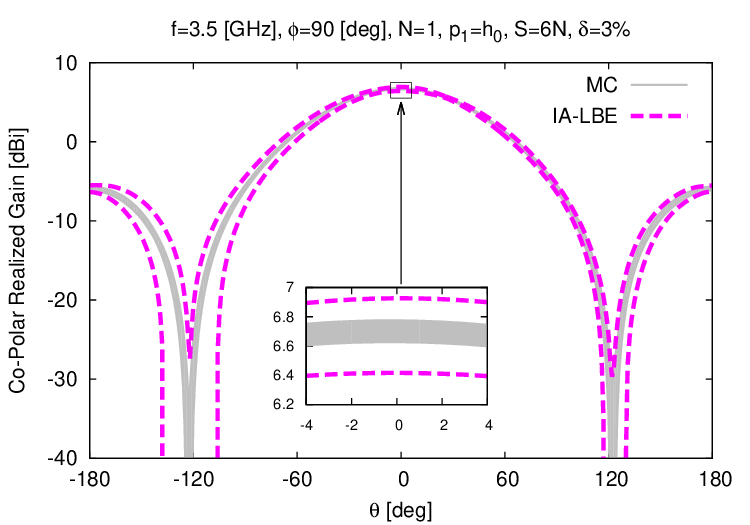}&
\includegraphics[%
  width=0.33\columnwidth]{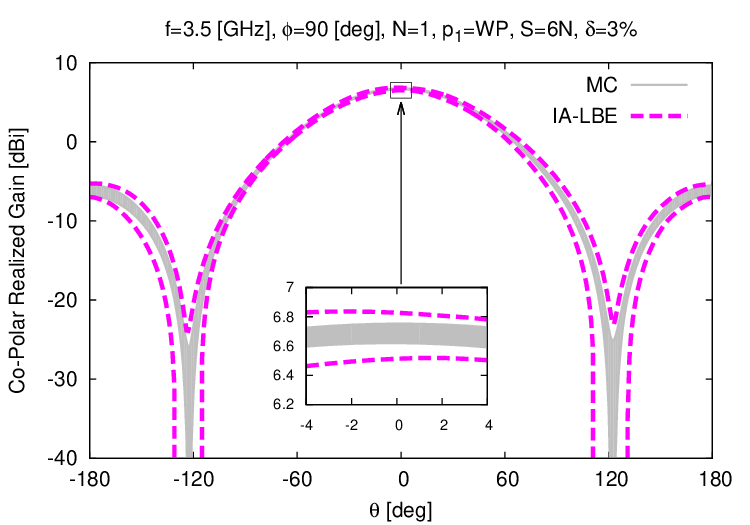}&
\includegraphics[%
  width=0.33\columnwidth]{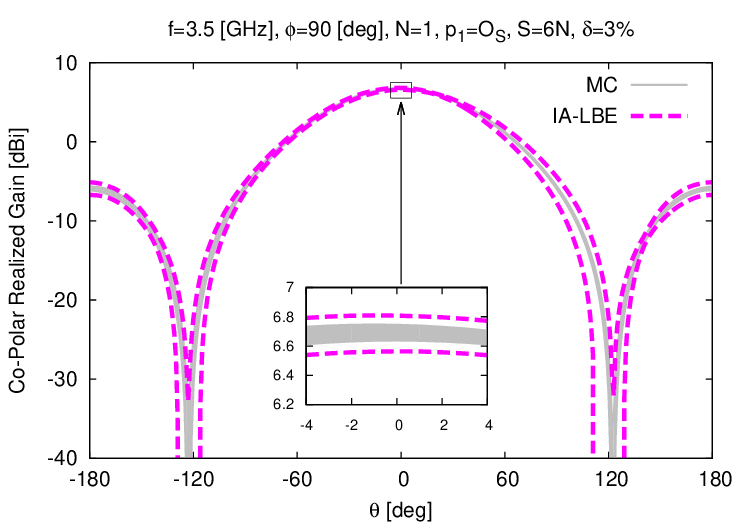}&
\includegraphics[%
  width=0.33\columnwidth]{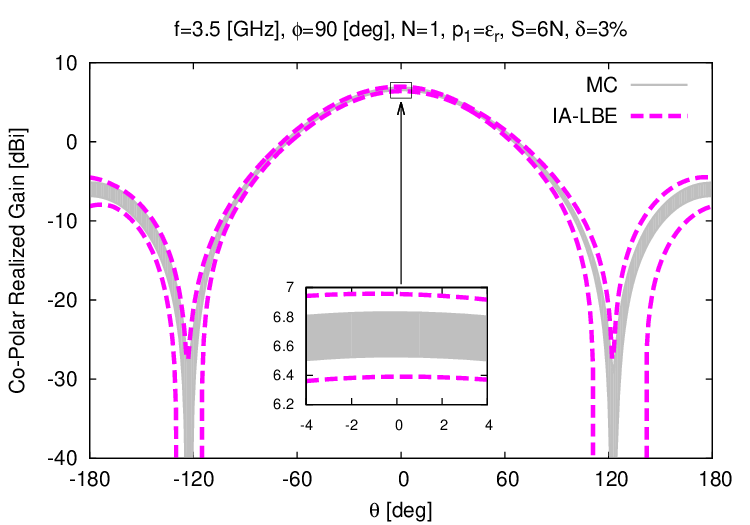}\tabularnewline
(\emph{e})&
(\emph{f})&
(\emph{g})&
(\emph{h})\tabularnewline
\end{tabular}
\end{sideways}\end{center}

\begin{center}\textbf{Fig. 15 - N. Anselmi} \textbf{\emph{et al.}}\textbf{,}
\textbf{\emph{{}``}}Sensitivity Analysis for Antenna Devices ...''\end{center}

\newpage
\begin{center}~\vfill\end{center}

\begin{center}\begin{tabular}{c}
\includegraphics[%
  width=0.70\columnwidth]{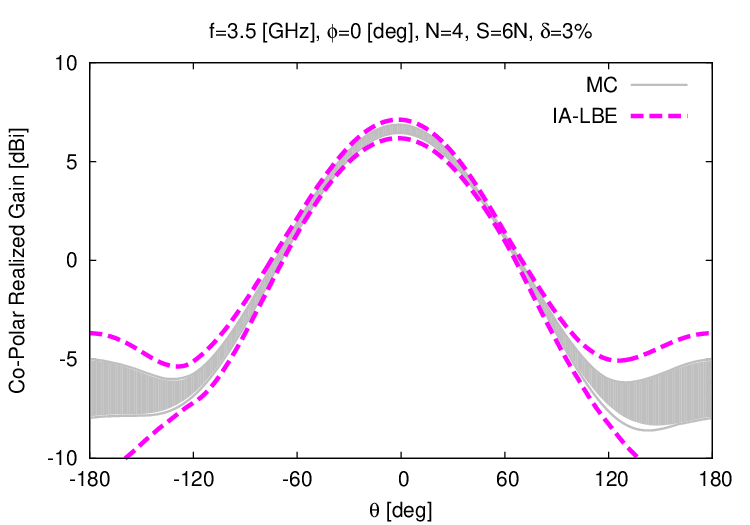}\tabularnewline
(\emph{a})\tabularnewline
\includegraphics[%
  width=0.70\columnwidth]{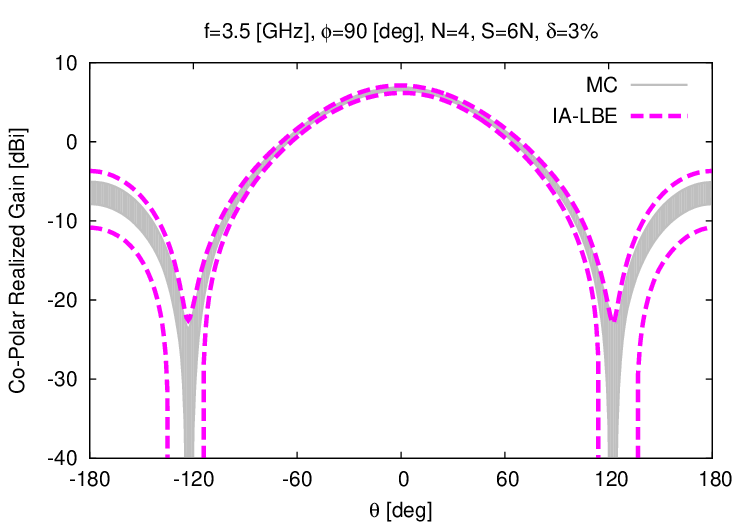}\tabularnewline
(\emph{b})\tabularnewline
\end{tabular}\end{center}

\begin{center}~\vfill\end{center}

\begin{center}\textbf{Fig. 16 - N. Anselmi} \textbf{\emph{et al.}}\textbf{,}
\textbf{\emph{{}``}}Sensitivity Analysis for Antenna Devices ...''\end{center}

\newpage
\begin{center}~\vfill\end{center}

\begin{center}\begin{tabular}{|c||c|c|c||c|}
\hline 
$\frac{S}{N}$&
$\Psi_{int}$ &
$\Psi_{ext}$&
$\Psi_{pen}$&
$\Psi$\tabularnewline
\hline
\hline 
$1$&
$1.01$&
$5.05\times10^{-3}$&
$5.05\times10^{-1}$&
$-1.51$\tabularnewline
\hline 
$2$&
$9.38\times10^{-1}$&
$0.00$&
$0.00$&
$-9.38\times10^{-1}$\tabularnewline
\hline
$3$&
$6.26\times10^{-1}$&
$0.00$&
$0.00$&
$-6.26\times10^{-1}$\tabularnewline
\hline
$4$&
$3.99\times10^{-1}$&
$0.00$&
$0.00$&
$-3.99\times10^{-1}$\tabularnewline
\hline
$5$&
$3.10\times10^{-1}$&
$0.00$&
$0.00$&
$-3.10\times10^{-1}$\tabularnewline
\hline
$6$&
$0.00$&
$8.35\times10^{-2}$&
$0.00$&
$8.35\times10^{-2}$\tabularnewline
\hline
$7$&
$0.00$&
$1.74\times10^{-1}$&
$0.00$&
$1.74\times10^{-1}$\tabularnewline
\hline
$8$&
$0.00$&
$2.46\times10^{-1}$&
$0.00$&
$2.46\times10^{-1}$\tabularnewline
\hline
\end{tabular}\end{center}

\begin{center}~\vfill\end{center}

\begin{center}\textbf{Tab. I - N. Anselmi} \textbf{\emph{et al.}}\textbf{,}
\textbf{\emph{{}``}}Sensitivity Analysis for Antenna Devices ...''\end{center}

\newpage
\begin{center}~\vfill\end{center}

\begin{center}\begin{tabular}{|c||c|c|c|c|c||c|}
\hline 
&
$\delta$ {[}\%{]}&
$SLL$~{[}dB{]}&
$BW$~{[}u{]}&
$\Phi_{max}$~{[}dB{]}&
$\Delta$&
$\Psi$\tabularnewline
\hline
\hline 
\emph{Nominal}&
-&
$-20.00$&
$0.20$&
$17.92$&
-&
\tabularnewline
\hline
\hline 
&
$1$&
$\left[-20.65;\,-19.20\right]$&
$\left[0.19;\,0.21\right]$&
$\left[17.83;\,18.01\right]$&
$0.27$&
$0.82$\tabularnewline
\cline{2-2} \cline{3-3} \cline{4-4} \cline{5-5} \cline{6-6} \cline{7-7} 
\multicolumn{1}{|c||}{\emph{IA-E}}&
$5$&
$\left[-23.54;\,-16.44\right]$&
$\left[0.16;\,0.22\right]$&
$\left[17.47;\,18.34\right]$&
$1.36$&
$0.78$\tabularnewline
\cline{2-2} \cline{3-3} \cline{4-4} \cline{5-5} \cline{6-6} \cline{7-7} 
&
$10$&
$\left[-28.56;\,-13.64\right]$&
$\left[0.13;\,0.25\right]$&
$\left[17.01;\,18.75\right]$&
$2.79$&
$0.71$\tabularnewline
\hline
\hline 
&
$1$&
$\left[-21.14;\,-18.79\right]$&
$\left[0.19;\,0.21\right]$&
$\left[17.76;\,18.08\right]$&
$0.43$&
$1.61$\tabularnewline
\cline{2-2} \cline{3-3} \cline{4-4} \cline{5-5} \cline{6-6} \cline{7-7} 
\multicolumn{1}{|c||}{\emph{IA-LBE}}&
$5$&
$\left[-27.24;\,-14.84\right]$&
$\left[0.14;\,0.24\right]$&
$\left[17.06;\,18.70\right]$&
$2.23$&
$1.54$\tabularnewline
\cline{2-2} \cline{3-3} \cline{4-4} \cline{5-5} \cline{6-6} \cline{7-7} 
\multicolumn{1}{|c||}{}&
$10$&
$\left[-53.79;\,-11.10\right]$&
$\left[0.10;\,0.28\right]$&
$\left[16.13;\,19.39\right]$&
$4.51$&
$1.29$\tabularnewline
\hline
\end{tabular}\end{center}

\begin{center}~\vfill\end{center}

\begin{center}\textbf{Tab. II - N. Anselmi} \textbf{\emph{et al.}}\textbf{,}
\textbf{\emph{{}``}}Sensitivity Analysis for Antenna Devices ...''\end{center}

\newpage
\begin{center}~\vfill\end{center}

\begin{center}\begin{tabular}{|c||c|c|c|c|c||c|}
\hline 
&
$\delta$ {[}\%{]}&
$SLL$~{[}dB{]}&
$BW$~{[}u{]}&
$\Phi_{max}$~{[}dB{]}&
$\Delta$&
$\Psi$\tabularnewline
\hline
\hline 
\multicolumn{6}{|c||}{$N=20$}&
\tabularnewline
\hline
\hline 
\emph{Nominal}&
-&
$-20.00$&
$0.01$&
$20.00$&
-&
-\tabularnewline
\hline
\cline{2-2} \cline{3-3} \cline{4-4} \cline{5-5} \cline{6-6} \cline{7-7} 
\multicolumn{1}{|c||}{\emph{IA-E}}&
$10$&
$\left[-28.70;\,-13.58\right]$&
$\left[0.06;\,0.12\right]$&
$\left[19.08;\,20.83\right]$&
$3.57$&
$1.21$\tabularnewline
\cline{2-2} \cline{3-3} \cline{4-4} \cline{5-5} \cline{6-6} \cline{7-7} 
\hline 
\multicolumn{1}{|c||}{\emph{IA-LBE}}&
$10$&
$\left[-45.30;\,-11.34\right]$&
$\left[0.00;\,0.14\right]$&
$\left[18.28;\,21.43\right]$&
$5.49$&
$1.89$\tabularnewline
\hline
\hline 
\multicolumn{6}{|c||}{$N=50$}&
\tabularnewline
\hline
\hline 
\multicolumn{1}{|c||}{\emph{Nominal}}&
-&
$-20.00$&
$0.04$&
$25.23$&
-&
-\tabularnewline
\hline
\cline{1-1} 
\multicolumn{1}{|c||}{\emph{IA-E}}&
$10$&
$\left[-28.83;\,-13.57\right]$&
$\left[0.02;\,0.05\right]$&
$\left[24.31;\,26.05\right]$&
$6.58$&
$2.22$\tabularnewline
\hline
\cline{1-1} 
\multicolumn{1}{|c||}{\emph{IA-LBE}}&
$10$&
$\left[-49.19;\,-11.04\right]$&
$\left[0.00;\,0.05\right]$&
$\left[23.52;\,26.65\right]$&
$9.79$&
$3.10$\tabularnewline
\hline
\end{tabular}\end{center}

\begin{center}~\vfill\end{center}

\begin{center}\textbf{Tab. III - N. Anselmi} \textbf{\emph{et al.}}\textbf{,}
\textbf{\emph{{}``}}Sensitivity Analysis for Antenna Devices ...''\end{center}

\newpage
\begin{center}~\vfill\end{center}

\begin{center}\begin{tabular}{|c|c|}
\hline 
\emph{Parameter}&
\emph{Value}\tabularnewline
\hline
\hline 
$h_{0}$&
$7.60\times10^{-4}$ {[}m{]}\tabularnewline
\hline 
$h_{1}$&
$4.56\times10^{-3}$ {[}m{]}\tabularnewline
\hline 
$h_{2}$&
$4.56\times10^{-3}$ {[}m{]}\tabularnewline
\hline 
$h_{p}=h_{G}$&
$3.50\times10^{-5}$ {[}m{]}\tabularnewline
\hline 
$LP=WP$&
$2.04\times10^{-2}$ {[}m{]}\tabularnewline
\hline 
$WF$&
$4.08\times10^{-3}$ {[}m{]}\tabularnewline
\hline 
$L_{F}$&
$4.66\times10^{-3}$ {[}m{]}\tabularnewline
\hline 
$L_{T}$&
$3.51\times10^{-3}$ {[}m{]}\tabularnewline
\hline 
$O_{S}$&
$1.60\times10^{-3}$ {[}m{]}\tabularnewline
\hline 
$WS$&
$5.94\times10^{-3}$ {[}m{]}\tabularnewline
\hline 
$LS$&
$2.04\times10^{-2}$ {[}m{]}\tabularnewline
\hline 
$\varepsilon_{r}$&
$3.0$\tabularnewline
\hline
$\tan\delta$&
$1.6\times10^{-3}$\tabularnewline
\hline
\end{tabular}\end{center}

\begin{center}~\vfill\end{center}

\begin{center}\textbf{Tab. IV - N. Anselmi} \textbf{\emph{et al.}}\textbf{,}
\textbf{\emph{{}``}}Sensitivity Analysis for Antenna Devices ...''\end{center}

\newpage
\begin{center}~\vfill\end{center}

\begin{center}\begin{tabular}{|c|c|c|c|}
\hline 
\multicolumn{1}{|c|}{$p_{n}$}&
$\delta$ {[}\%{]}&
$\left|\frac{\inf}{\sup}\left(\delta_{n}\right)\right|$~{[}mm{]}&
\multicolumn{1}{c|}{$\left[p_{n}\right]$~{[}mm{]}}\tabularnewline
\hline
\hline 
$h_{0}$&
$3$&
$0.0228$ &
$\left[0.7372;\,0.7828\right]$\tabularnewline
\hline
$WP$&
$3$&
$0.6120$&
$\left[19.7880;\,21.0120\right]$\tabularnewline
\hline
\multicolumn{1}{|c|}{$O_{S}$}&
$3$&
$0.0480$&
$\left[1.5520;\,1.6480\right]$\tabularnewline
\hline
\cline{1-1} 
\multicolumn{1}{|c|}{$\epsilon_{r}$}&
$3$&
$0.0900$&
$\left[2.9100;\,3.0900\right]$\tabularnewline
\hline
\end{tabular}\end{center}

\begin{center}~\vfill\end{center}

\begin{center}\textbf{Tab. V - N. Anselmi} \textbf{\emph{et al.}}\textbf{,}
\textbf{\emph{{}``}}Sensitivity Analysis for Antenna Devices ...''\end{center}

\newpage
\begin{center}~\vfill\end{center}

\begin{center}\begin{tabular}{|c||c|c|c|}
\hline 
$N$ &
$\underline{p}$&
\multicolumn{2}{c|}{$\Psi_{IA-LBE}$}\tabularnewline
\hline
\hline 
-&
-&
$\phi=0$ {[}deg{]}&
$\phi=90$ {[}deg{]}\tabularnewline
\hline
\hline 
$1$ &
 $h_{0}$&
$1.55$&
$2.66$\tabularnewline
\hline 
\multicolumn{1}{|c||}{$1$ }&
\multicolumn{1}{c|}{ $WP$}&
$1.02$&
$1.66$\tabularnewline
\cline{3-3} \cline{4-4} 
\hline 
\multicolumn{1}{|c||}{$1$ }&
\multicolumn{1}{c|}{$O_{S}$}&
$1.34$&
$2.38$\tabularnewline
\cline{3-3} \cline{4-4} 
\hline 
\multicolumn{1}{|c||}{$1$ }&
\multicolumn{1}{c|}{$\epsilon_{r}$}&
$0.90$&
$1.01$\tabularnewline
\cline{3-3} \cline{4-4} 
\hline 
\multicolumn{1}{|c||}{$4$}&
\multicolumn{1}{c|}{$\left\{ h_{0},\, WP,\, O_{S},\,\epsilon_{r}\right\} $ }&
$1.51$&
$2.45$\tabularnewline
\hline
\end{tabular}\end{center}

\begin{center}~\vfill\end{center}

\begin{center}\textbf{Tab. VI - N. Anselmi} \textbf{\emph{et al.}}\textbf{,}
\textbf{\emph{{}``}}Sensitivity Analysis for Antenna Devices ...''\end{center}
\end{document}